\documentclass[10pt,final,doublecolumn]{IEEEtran}
\usepackage[caption=false,font=normalsize,labelfont=sf,textfont=sf]{subfig}
\usepackage{titlesec}
\titlespacing*{\subsection}{0pt}{0.5\baselineskip}{0.3\baselineskip}

\usepackage{amsmath}
\usepackage{amssymb}   
\usepackage{amsthm}
\usepackage{graphicx}
\usepackage{bbding}
\usepackage{indentfirst}
\usepackage{algorithm,algorithmic}
\usepackage{setspace}
\usepackage{float}
\usepackage{amsfonts}
\usepackage{enumerate}
\usepackage{multicol}
\usepackage{color}
\usepackage{bm}
\usepackage{stfloats}
\usepackage{epstopdf}
\usepackage{threeparttable}
\usepackage{makecell}
\usepackage{cite}
\usepackage{booktabs} 
\usepackage{ulem}

\newcommand{\mbf}[1]{\mathbf{#1}}
\newcommand{\mbb}[1]{\mathbb{#1}}
\newcommand{\mc}[1]{\mathcal{#1}}
\newcommand{\expa}[1]{\mathbb{E}\left[ #1 \right]}

\newcommand{\trace}[1]{\text{tr}\left(#1\right)}

\theoremstyle{remark}
\newtheorem{remark}{Remark}

\begin{document}
\title{ Statistical Channel Based Low-Complexity  Rotation and Position Optimization for 6D Movable Antennas  Enabled Wireless Communication }

\author{
	Qijun Jiang, 
	Xiaodan Shao,~\IEEEmembership{Member,~IEEE}, 
	and Rui Zhang,~\IEEEmembership{Fellow,~IEEE}
\thanks{This paper was partially submitted to IEEE/CIC International Conference on Communications in China (ICCC) Workshop, 2025 \cite{conf6}.}
\thanks{Q. Jiang is with the School of Science and Engineering, The Chinese University of Hong Kong, Shenzhen, Guangdong 518172, China (e-mail: qijunjiang@link.cuhk.edu.cn).}
\thanks{X. Shao is with the Department of Electrical and Computer Engineering, University of Waterloo, Waterloo, ON N2L 3G1, Canada (e-mail: x6shao@uwaterloo.ca).}
\thanks{R. Zhang is with the School of Science and Engineering, Shenzhen Research Institute of Big Data, The Chinese University of Hong Kong, Shenzhen, Guangdong 518172, China. He is also with the Department of Electrical and Computer Engineering, National University of Singapore, Singapore 117583 (e-mail: elezhang@nus.edu.sg).}
}

\maketitle

\begin{abstract}
	Six-dimensional movable antenna (6DMA) is a promising technology to fully exploit spatial variation in wireless channels by allowing flexible adjustment of three-dimensional (3D) positions and rotations of antennas at the transceiver. In this paper, we investigate the practical low-complexity design of 6DMA-enabled communication systems, including transmission protocol, statistical channel information (SCI) acquisition, and joint position and rotation optimization of 6DMA surfaces based on the SCI of users. Specifically, an orthogonal matching pursuit (OMP)-based algorithm is proposed for the estimation of SCI of users at all possible position-rotation pairs of 6DMA surfaces based on the channel measurements at a small subset of position-rotation pairs.  
	Then, the average sum logarithmic rate of all users  is maximized by jointly designing the positions and rotations of 6DMA surfaces based on their SCI acquired. Different from prior works on 6DMA which adopt alternating optimization to design 6DMA positions/rotations with iterations, we propose a new sequential optimization approach that first determines 6DMA rotations and then finds their feasible positions to realize the optimized rotations subject to practical antenna placement constraints. Simulation results show that the proposed sequential optimization significantly reduces the computational complexity of conventional alternating optimization, while achieving comparable communication performance. It is also shown that the proposed SCI-based 6DMA design can effectively enhance the communication throughput of wireless networks over existing fixed (position and rotation) antenna arrays, yet with a practically appealing low-complexity implementation.            
\end{abstract}


\section{Introduction}
Over the past decades, numerous technological advancements have contributed to the development of mobile communication networks. Among them, multi-antenna or multiple-input multiple-output (MIMO) technology is widely regarded as the most significant propeller for the past generations of wireless communication systems. The recent evolution of MIMO technologies, including cell-free massive MIMO \cite{RN51}, extremely large-scale MIMO \cite{LargeMIMO, exl}, and intelligent reflecting surface (IRS)-aided MIMO \cite{IRS,huang2019RIS,add_1}, has further enhanced the MIMO performance gains in wireless networks by deploying increasingly more active and/or passive antennas. However, in  conventional MIMO systems,  antennas are usually placed  at fixed, equally spaced positions, which limits their ability to fully utilize the spatial variation of wireless channels at the transmit and receiver \cite{zhu2024movable}.

To address this issue, six-dimensional movable antenna (6DMA)  has recently emerged as a promising solution \cite{6DMA_TWC}. 6DMA provides the highest flexibility to exploit the spatial channel variation at wireless transceivers  by allowing independent adjustment of both the three-dimensional (3D) positions and rotations of antennas or antenna surfaces, with practically affordable slow adaptation  based on the spatial channel distribution of users in the network. Prior studies have demonstrated the effectiveness of 6DMA in significantly enhancing the communication and sensing performance of wireless networks by exploiting the new joint 6DMA position and rotation optimization  \cite{6DMA_TWC, 6DMA_JSAC,6DMA_Xiaoming, 6DMA_JSTSP, add_3, passive6DMA, UAV6DMA ,  free6DMA, 6dmasensing,add_2}. Specifically, the work  \cite{6DMA_TWC} introduces the 6DMA architecture and formulates the ergodic capacity maximization problem, demonstrating the significant network capacity improvement obtained through continuous adjustment of 6DMA surface rotations and positions over the conventional  fixed (position and rotation)  antenna arrays. In \cite{6DMA_JSAC}, the optimization problem for 6DMA is extended to the case with practical discrete position and rotation levels, where a new online learning and optimization approach is proposed without any prior knowledge of user channel 
distribution. In addition, \cite{6DMA_Xiaoming} proposes a hybrid base station (BS) architecture that integrates fixed antenna arrays and 6DMA surfaces for performance enhancement. Moreover, the authors in \cite{6DMA_JSTSP} propose channel estimation algorithms for both instantaneous and statistical channels by exploiting a new directional sparsity characteristic of 6DMA channels for training complexity reduction. The polarized 6DMA (also called a polarforming antenna) is proposed in \cite{add_3}, which can enable polarforming to adaptively control the antenna’s polarization electrically and to tune its position and rotation mechanically, thereby effectively exploiting polarization and spatial diversity to improve system performance.   Furthermore, the studies in \cite{passive6DMA, UAV6DMA} investigate MIMO systems assisted by passive 6DMA, while \cite{free6DMA} examines 6DMA-aided cell-free networks. Besides communication, the authors in \cite{6dmasensing} propose a wireless sensing system that incorporates 6DMA. In the above works, the performance advantages of 6DMA over antenna position-adjustable  fluid antennas or movable antennas \cite{MA, FAS, Zhang2025, Hu2024} are demonstrated under various system setups.

Despite the great potential of 6DMA for wireless communication and sensing, its practical deployment in wireless networks faces new challenges. First, low-complexity optimization of positions and rotations of a given number of 6DMA surfaces in a large region (or a large number of discretized position-rotation pairs therein) is a crucial but  challenging problem. Nevertheless, 6DMA optimization requires only the statistical channel information (SCI) of users in the network, instead of their instantaneous channels.   This is because the SCI of users  depends on their   spatial distribution as well as the dominant  scatterers in the environment, thus changing much more slowly as compared to the users' instantaneous channels.  
As the movement speed of 6DMA surfaces is practically constrained by mechanical drivers, their position/rotation adaptation based on users' SCI can significantly reduce the surface movement frequency as compared to that for adapting to users' instantaneous channels\cite{MA,FAS}, thus making 6DMA more practically implementable.     
However, the existing works on 6DMA position and rotation optimization (e.g.,  \cite{6DMA_TWC,6DMA_Xiaoming}) assume \textit{a priori} known SCI of users and apply the Monte Carlo method to generate a large set of users' instantaneous channels based on their known SCI, thereby approximating the network ergodic capacity by averaging over the generated channel samples. This approach not only incurs high computational complexity, but its accuracy also relies on the size of the channel sample set, which needs to increase with the number of users and the number of antennas at the 6DMA-equipped BS. Moreover, the optimization proposed in \cite{6DMA_TWC} is based on alternating optimization, which iteratively designs 6DMA positions/rotations with the other being fixed, which may converge to     undesired local optima, thus requiring a properly designed initialization method to ensure the performance of the converged solution.  

Second, the SCI of all users in the entire 6DMA movement region (or all discretized 6DMA position-rotation pairs in it)  is essential for optimizing antenna positions and rotations. However, different from  conventional fixed antenna arrays  \cite{LargeMIMO, exl} for which the channel dimension is fixed for both SCI and instantaneous channel estimations, how to estimate the SCI of all users at all possible position-rotation pairs for 6DMA surfaces in a given region based on the channel measurements at a finite number (small subset) of position-rotation pairs is a new and challenging problem that remains unsolved.

To  tackle the aforementioned design  challenges for 6DMA, this paper investigates a practical low-complexity design of 6DMA-enabled communication systems, including transmission protocol, SCI acquisition, and joint rotation and position optimization of 6DMA surfaces based on the acquired SCI of users.
The main contributions of this paper are summarized as follows.
\begin{itemize}
	\item
	First, we propose a three-stage protocol for  6DMA-enabled wireless communication systems. In the first stage, the 6DMA surfaces move to selected training position and rotation pairs within the movement region to collect channel measurements required for the estimation of users' SCI in the entire region. In the second stage, the SCI of all users is reconstructed, based on which the positions and rotations of all 6DMA surfaces are optimized, and then they are moved to the designed positions and rotations. In the third stage,  the 6DMA-equipped BS  serves all users with enhanced communication performance.

	\item Second, we present an orthogonal matching pursuit (OMP)-based method for the SCI reconstruction. Specifically, the multi-path component (MPC) information of each user, including the direction of arrivals (DOAs) and the average power values of all different paths, is first estimated based on the channel measurements obtained at the 6DMA training position-rotation pairs  using the OMP algorithm. Then, the SCI of all users is reconstructed at any position and rotation in the 6DMA movement region. {Moreover, we analyze the effects of key system parameters on  the accuracy of SCI estimation, including the number of training antenna positions and
		rotations, as well as the  beamwidth of directional antennas used at the 6DMA-BS.}
	
	\item
	Third, we formulate an optimization problem for designing 6DMA  positions and rotations to maximize the  average sum
	logarithmic rate (log-rate) of a multi-user (uplink) multiple-access channel (MAC) by considering the rate fairness among users, based on their estimated SCI. To reduce the computational complexity of the conventional Monte Carlo method, we derive an analytical approximation for the  average achievable 
	rates of users in terms of the 6DMA positions and rotations. Furthermore, to avoid the  undesired local optima of the existing alternating optimization approach with iterations, we propose a new sequential optimization approach that first determines 6DMA rotations and then finds their feasible positions to realize the optimized rotations subject to practical antenna placement constraints.
	
	\item
	Finally, simulation results are presented which validate  the effectiveness of the proposed SCI acquisition method with low training overhead, as well as our proposed low-complexity  sequential 6DMA rotation and position  optimization algorithm based on SCI, which achieves performance close to the  Monte Carlo and alternating optimization based  optimization,   while significantly reducing the computational complexity.
\end{itemize}

\newcounter{nnn} 
\setcounter{nnn}{2} 
The rest of this paper is organized as follows. Section \Roman{nnn}\addtocounter{nnn}{1} presents the system model and problem formulation. Section \Roman{nnn}\addtocounter{nnn}{1} proposes a practical protocol for 6DMA-BS. Section \Roman{nnn}\addtocounter{nnn}{1} introduces the SCI estimation algorithm. Section \Roman{nnn}\addtocounter{nnn}{1} presents the proposed algorithm for optimizing 6DMA surfaces' rotations and positions sequentially based on SCI.  Section \Roman{nnn}\addtocounter{nnn}{1} provides simulation results for performance evaluation and comparison. Finally, Section \Roman{nnn}\addtocounter{nnn}{1} concludes this paper.

\textit{Notations}: Symbols for vectors (lower case) and matrices (upper case) are in boldface. Symbols for sets are denoted using calligraphic letters. $(\cdot)^\top$ , $(\cdot)^\dag$, and $(\cdot)^H$ denote the transpose, conjugate, and conjugate transpose (Hermitian) operations, respectively. $\lfloor\cdot\rfloor$ and $\lceil\cdot\rceil$ denote the floor and ceiling operations, respectively.  The sets of $P\times Q$ dimensional complex and real matrices are denoted by $\mbb{C}^{P\times Q}$ and $\mbb{R}^{P\times Q}$, respectively.  The sets of $P$-dimensional complex and real vectors are denoted by $\mbb{C}^{P}$ and $\mbb{R}^{P}$, respectively. $\mbb{E}(\cdot)$ denotes the  expectation operator. $\otimes$ denotes Kronecker product. We use $\text{diag}(a_1,\cdots,a_P)$ to denote the square diagonal matrix with entries $a_1,\cdots,a_P$ on its main diagonal. $\|\mbf{a}\|_2$ and $\|\mbf{A}\|_F$ denote the $l_2$ norm of vector $\mbf{a}$ and Frobenius norm of matrix $\mbf{A}$, respectively. The vectorization of matrix $\mbf{A}$ is denoted by $\mathrm{vec}(\mbf{A})$.

 \section{System Model and Problem Formulation}
\subsection{Channel Model}
\begin{figure}[!t]  
    \centering
    \includegraphics[width=2.5in]{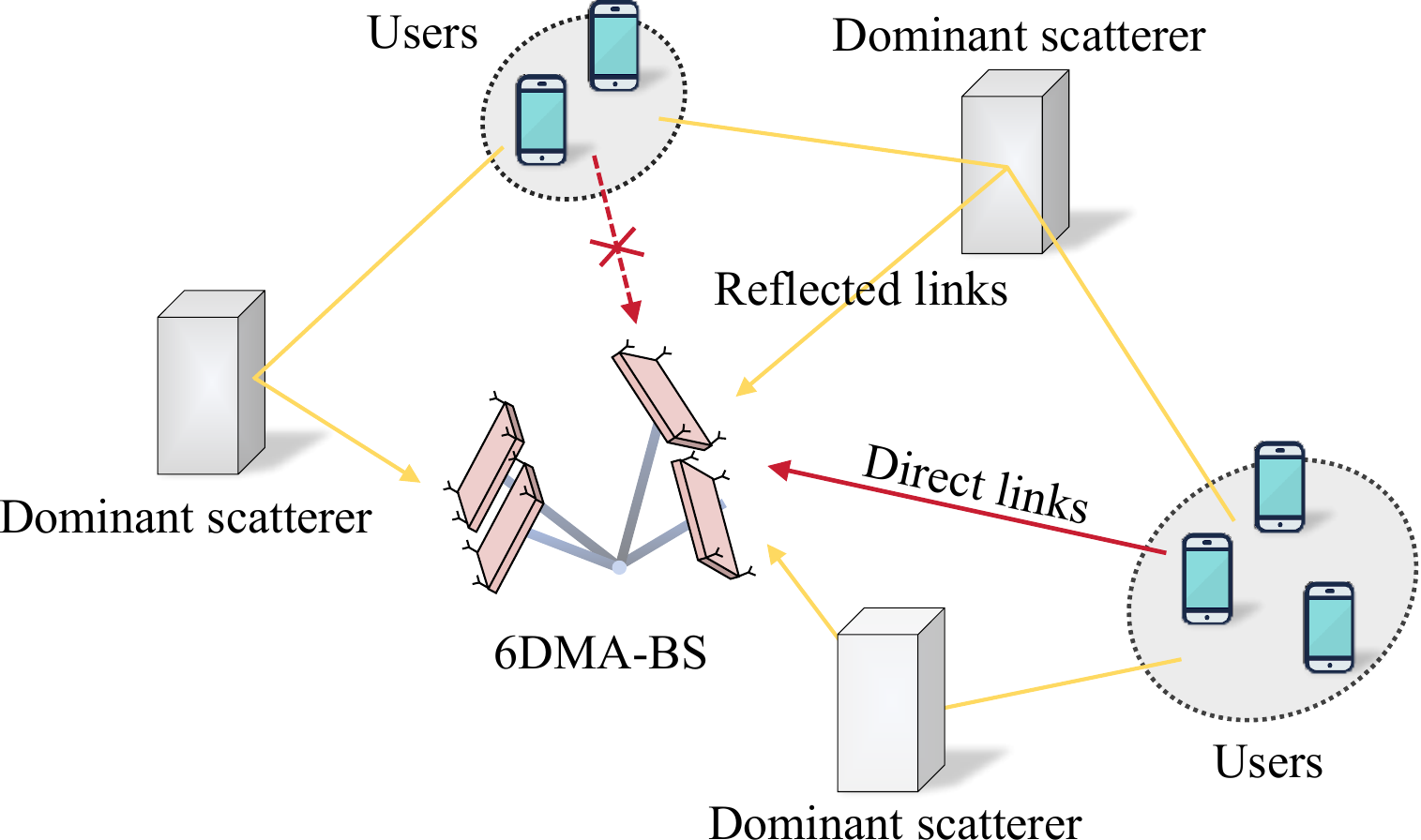}  
    \caption{A 6DMA-enabled wireless communication system.}  
    \label{fig:Channel_model}  
\end{figure}

\begin{figure}[!t]  
	\centering
	\includegraphics[width=1.4in]{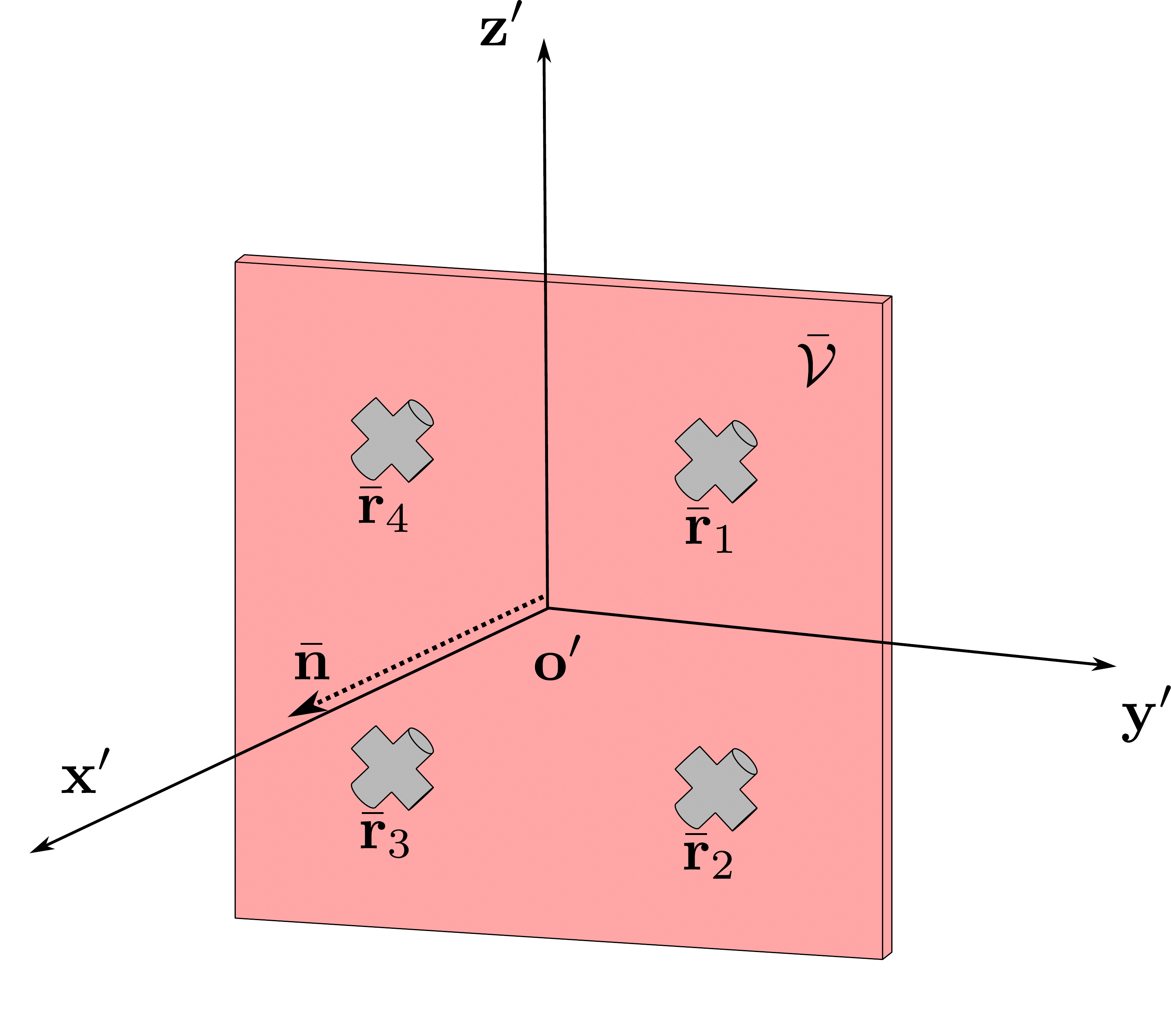}  
	\caption{An example of  6DMA surface geometry. The  square region  of one 6DMA surface, $\mc{\bar{V}}$, is highlighted in pink. The local CCS origin of the surface, $o'$, is located at the square's geometric center. The normal vector of the square, $\bar{\mbf{n}}$, which  coincides with the direction of the antenna main lobe, is aligned with the ${o}'$-${x}'$ axis of the local CCS. The edges of the square are parallel to the ${o}'$-${y}'$  or  ${o}'$-${z}'$ axes. In total, $N=4$ antennas are placed within the square region, with their positions  represented by $\bar{\mbf{r}}_1,\cdots,\bar{\mbf{r}}_4$ respectively in the local CCS.}  
	\label{fig:single_surface}  
\end{figure}

We consider a 6DMA-aided  communication system with one single 6DMA-BS serving $K$ distributed users. As shown in Fig.\,\ref{fig:Channel_model},
each user $k$, $k \in \mc{K} \triangleq \{1,\cdots,K\}$, has  $L_{k}$ propagation  paths to the BS, indexed by the set $\mc{L}_k \triangleq \{1,\cdots,L_k\}$, which result in the direct link and/or  reflected links  by dominant scatterers. The 6DMA-BS is equipped with $B$ 6DMA surfaces, indexed by the set $\mc{B} \triangleq \{1,\cdots,B\}$,  each  comprising $N$  directional antennas\footnote{This is to be consistent with practical BS antenna models\cite{3gpp2017}.}, indexed by the set $\mc{N}\triangleq\{1,\cdots,N\}$. All users are each equipped with an omni-directional antenna. The positions and rotations (orientations) of all 6DMA surfaces can be individually adjusted. In particular, the position and rotation  of the $b$-th 6DMA surface, $b\in \mc{B}$, can be respectively characterized by
  \begin{align}
\mbf{q}_b &\triangleq [x_b,y_b,z_b]^\top \in \mc{V}_{\text{6DMA}},\\
\mbf{u}_b &\triangleq [\alpha_b, \beta_b, \gamma_b]^\top,
\end{align}
where $\mathcal{V}_{\text{6DMA}}$ denotes the given 3D region at the BS in which the 6DMA surfaces can be flexibly positioned/rotated. In the above, $x_b$, $y_b$, and $z_b$ represent the coordinates of the center of the $b$-th 6DMA surface in the global Cartesian coordinate system (CCS) $o$-$xyz$, with the 6DMA-BS’s reference position serving as the origin $o$; $\alpha_b $, $\beta_b $, and $\gamma_b $  denote the Euler angles with respect to (w.r.t.) the $x$-axis, $y$-axis and $z$-axis in the global CCS, respectively, in the range of $\left[0,2\pi\right)$.  We define the position-rotation pair $\mbf{z}_b \triangleq  [\mbf{q}_b^\top, \mbf{u}_b^\top]^\top \in \mbb{R}^{6}$ to  compactly represent the position and rotation of the $b$-th 6DMA surface. The end-to-end channels between the $K$ users and the 6DMA-BS as a function of the position-rotation state $\mbf{z} \triangleq [\mbf{z}_1^\top,\cdots,\mbf{z}_B^\top]^\top \in \mbb{R}^{6 B}$ of all  6DMA surfaces can be expressed as
\begin{align}
\mbf{H}(\mbf{z}) \triangleq \begin{bmatrix} \mbf{h}_1(\mbf{z}_1) & \cdots & \mbf{h}_K(\mbf{z}_1)\\
                                           \vdots        &        & \vdots \\
                                           \mbf{h}_1(\mbf{z}_B) & \cdots & \mbf{h}_K(\mbf{z}_B)\\
                            \end{bmatrix}  \in \mbb{C}^{BN\times K},
\end{align}
where $\mbf{h}_k(\mbf{z}_b) \in \mathbb{C}^{N}$ denotes the channel from the $k$-th user to the $b$-th 6DMA surface with position-rotation pair $\mbf{z}_b$.  For the purpose of exposition, we consider a narrow-band system with flat-fading channels, thus $\mbf{h}_k(\mbf{z}_b) $ can be expressed as
\begin{align}
\mbf{h}_k(\mbf{z}_b) &= \sum_{l=1}^{L_k} \sqrt{g(\mbf{u}_b,\mbf{f}_{k,l})}\mathbf{{a}}(\mbf{z}_b,\mbf{f}_{k,l})  v_{k,l}, \label{hkb}
\end{align}
where 
$\mbf{f}_{k,l}\in \mbb{R}^{3}$ and $v_{k,l}$, $k \in \mc{K}$, $l \in \mc{L}_k$, denote the DoA and complex channel coefficient of the $l$-th path of the $k$-th user, respectively;  $g(\mbf{u}_b,\mbf{f}_{k,l})$ 
 and $\mathbf{{a}}(\mbf{z}_b,\mbf{f}_{k,l}) \in \mbb{C}^{N}$, $b \in \mc{B}$, respectively denote the  antenna gain and  steering vector of the $b$-th 6DMA surface. Specifically, the antenna gain $g(\mbf{u}_b,\mbf{f}_{k,l})$ depends on the rotation $\mbf{u}_b$ of the $b$-th 6DMA surface  and the DoA $\mbf{f}_{k,l}$, while the steering vector $\mathbf{{a}}(\mbf{z}_b,\mbf{f}_{k,l})$ is determined by the position-rotation pair $\mbf{z}_b$ of the $b$-th 6DMA surface and the DoA $\mbf{f}_{k,l}$. More specifically, we have
\begin{align}
g(\mbf{u}_b,\mbf{f}_{k,l}) &= G\left(\mbf{R}^{-1}(\mbf{u}_b) \mbf{f}_{k,l}\right), \label{antenna_gain}\\
\mathbf{{a}}(\mbf{z}_b,\mbf{f}_{k,l}) &= [e^{-j\frac{2\pi}{\lambda}\mbf{f}_{k,l}^{\top} \mbf{r}_{1}(\mbf{z}_b)},\cdots,e^{-j\frac{2\pi}{\lambda}\mbf{f}_{k,l}^{\top} \mbf{r}_{N}(\mbf{z}_b)}]^\top \label{steering_vec_def},
\end{align}
where $\mbf{R}(\mbf{u}_b)$ is the rotation matrix corresponding to rotation $\mbf{u}_b$, 
 which is given by
\begin{align} &\mbf{R}(\mbf{u}_b) = \\
&\begin{bmatrix}
	c_{\beta_b} c_{\alpha_b} & -c_{\beta_b} s_{\alpha_b} & s_{\beta_b} \\
	s_{\gamma_b} s_{\alpha_b} + c_{\alpha_b} c_{\gamma_b} s_{\beta_b} & c_{\alpha_b} c_{\gamma_b} - s_{\alpha_b} s_{\gamma_b} s_{\beta_b} & -c_{\beta_b} s_{\gamma_b} \\
	s_{\gamma_b} s_{\alpha_b} - c_{\gamma_b} c_{\alpha_b} s_{\beta_b} & c_{\alpha_b} s_{\gamma_b} + c_{\gamma_b} s_{\alpha_b} s_{\beta_b} & c_{\gamma_b} c_{\beta_b}
\end{bmatrix}, \nonumber
\end{align}
with $c_x \triangleq \cos(x)$ and $s_x \triangleq \sin(x)$ for notational brevity.
In \eqref{antenna_gain}, it can be shown that  $\mbf{R}^{-1}(\mbf{u}_b) \mbf{f}_{k,l} = \mbf{R}^\top(\mbf{u}_b) \mbf{f}_{k,l}$, which denotes the DoA $\mbf{f}_{k,l}$ after being projected to the  local CCS of the $b$-th 6DMA surface with rotation $\mbf{u}_b$; the function $G(\cdot)$ denotes the effective gain of antennas of each 6DMA surface in terms of the DoA in its local CCS (to be specified in Section VI  based on the practical antenna radiation pattern adopted). In \eqref{steering_vec_def}, $\lambda$ denotes the wavelength of the carrier wave, and $\mbf{r}_n(\mbf{z}_b) \in \mbb{R}^3$, $n\in\mc{N}$, represents the location of the $n$-th antenna on the $b$-th 6DMA surface with the position-rotation pair $\mbf{z}_b$ in the global CCS, i.e., 
\begin{align}
\mbf{r}_n(\mbf{z}_b) = \mbf{q}_b + \mbf{R}(\mbf{u}_b) \bar{\mbf{r}}_n, \label{comformal_trans}
\end{align}
where the position of the $n$-th antenna of a 6DMA surface in its local CCS, $\bar{\mbf{r}}_n$, is predefined  based on the  practical geometry of 6DMA surfaces (e.g., uniform planar array (UPA)), as shown in Fig.\,\ref{fig:single_surface}. To reduce signal coupling between adjacent antennas,  $\bar{\mbf{r}}_n$'s are chosen such that the minimum distance between any two antennas on a 6DMA surface is no less than $\frac{\lambda}{2}$.

Let $\mbf{h}_k(\mbf{z}) \triangleq [\mbf{h}_k(\mbf{z}_1)^H, \cdots, \mbf{h}_k(\mbf{z}_B)^H ]^H \in \mbb{C}^{BN}$ denote the channel  from the $k$-th user to the 6DMA-BS by considering all its $B$ 6DMA surfaces. Following (\ref{hkb}), we have
\begin{align}
\mbf{h}_k(\mbf{z}) &= \sum_{l=1}^{L_k} \mbf{{G}}(\mbf{z},\mbf{f}_{k,l}) \mbf{a}(\mbf{z},\mbf{f}_{k,l}) v_{k,l} \nonumber \\
&= \sum_{l=1}^{L_k} \tilde{\mbf{a}}(\mbf{z},\mbf{f}_{k,l}) v_{k,l}
= \tilde{\mbf{A}}(\mbf{z},\mbf{F}_k) \mbf{v}_k,\label{h2}
\end{align}
where $\mbf{G}(\mbf{z},\mbf{f}_{k,l}) \in \mbb{R}^{BN\times BN}$  is a diagonal matrix with its $(i,i)$-th entry representing the antenna gain of the $i$-th antenna, $i=1,\cdots,BN$, of all the $B$ 6DMA surfaces. Specifically, 
\begin{align}
 \mbf{{G}}(\mbf{z},\mbf{f}_{k,l}) = \mathrm{diag}\left(\sqrt{g(\mbf{u}_1,\mbf{f}_{k,l})},\cdots,\sqrt{g(\mbf{u}_B,\mbf{f}_{k,l})}\right) \otimes \mbf{I}_N,
\end{align}
where
 the Kronecker product is used due to the fact that all antennas on a 6DMA surface share the same antenna gain since their rotations are identical; $\mbf{a}(\mbf{z},\mbf{f}_{k,l}) \triangleq [\mathbf{{a}}(\mbf{z}_1,\mbf{f}_{k,l})^H, \cdots, \mathbf{{a}}(\mbf{z}_B,\mbf{f}_{k,l})^H]^H\in \mbb{C}^{BN}$ is the steering vector from the $l$-th path of the $k$-th user to the $B$ 6DMA surfaces, while
 \begin{align}
 \tilde{\mbf{a}}(\mbf{z},\mbf{f}_{k,l})  \triangleq \mbf{{G}}(\mbf{z},\mbf{f}_{k,l}) \mbf{a}(\mbf{z},\mbf{f}_{k,l})\in \mbb{C}^{BN} \label{weighted_steering_vector}
 \end{align}
  denotes the steering vector $\mbf{a}(\mbf{z},\mbf{f}_{k,l})$ weighted by the antenna gain of each 6DMA surface;  the matrix $\tilde{\mbf{A}}(\mbf{z},\mbf{F}_k) \in \mbb{C}^{BN\times L_k}$ resembles the weighted steering vectors of all paths of the $k$-th user, i.e.,
  \begin{align}
  	 \tilde{\mbf{A}}(\mbf{z},\mbf{F}_k)  \triangleq [\tilde{\mbf{a}}(\mbf{z},\mbf{f}_{k,1}),\cdots,\tilde{\mbf{a}}(\mbf{z},\mbf{f}_{k,L_k})], \label{weighted_steering_matrix}
  	\end{align}
  where
 $\mbf{F}_k  \triangleq [\mbf{f}_{k,1},\cdots,\mbf{f}_{k,L_k}]\in \mbb{R}^{3\times L_k}$, and the vector $\mbf{v}_k  \triangleq [v_{k,1},\cdots,v_{k,L_k}]^\top \in \mbb{C}^{L_k}$ collects the complex channel coefficients of all the $k$-th user's  paths.
 
  In this paper, we assume rich local scatterers around each user thus the channel coefficients of/among each user/different users  are uncorrelated. Thus, for convenience, we model all users' channel coefficients,   $\mbf{v}_k$'s, to be independent circularly symmetric complex Gaussian (CSCG) random variables,   i.e.,
 \begin{align}
 	\mbf{v}_k \sim \mc{CN}(0,\mbf{D}_k),
 \end{align}
where $\mbf{D}_k \triangleq \text{diag}(\alpha_{k,1}^2,\cdots,\alpha_{k,L_k}^2) \in \mbb{R}^{L_k\times L_k}$ with its $(l,l)$-th entry denoting the average power  of the $l$-th path of the $k$-th user to the BS.
In other words, the channels of users are independently sampled from 
\begin{align}
	\mbf{h}_k(\mbf{z})  \sim \mc{CN}(0,\mbf{\Sigma}_k(\mbf{z})), \label{hk_distri}
\end{align}
where the channel covariance matrix $\mbf{\Sigma}_k(\mbf{z})$ is given by
\begin{align}
	\mbf{\Sigma}_k(\mbf{z}) = \tilde{\mbf{A}}(\mbf{z},\mbf{F}_k) \mbf{D}_k \tilde{\mbf{A}}(\mbf{z},\mbf{F}_k)^H. \label{R_def}
\end{align}
As a consequence of (\ref{hk_distri}) and (\ref{R_def}),  the statistical characteristics of all channels are fully determined by the position-rotation state of 6DMA-BS $\mbf{z}$ and MPC information including  the DoAs of users, $\mbf{F}_k$'s, and the average power values of all paths of users, $\mbf{D}_k$'s. For this reason, we define $\mbf{\Sigma}(\mbf{z}) = [\mbf{\Sigma}_1(\mbf{z}),\cdots,\mbf{\Sigma}_k(\mbf{z})] \in \mbb{C}^{BN\times BNK}$  as the users' SCI in this paper.

\subsection{Unified Constraint for Blockage and Overlap Avoidance}
In this subsection, we introduce the unified constraint for 6DMA surfaces' placement, which prevents mutual signal reflections and physical overlap between any two 6DMA surfaces. First, we define the non-positive halfspace associated with the $b$-th 6DMA surface as
\begin{align}
	\mc{H}^-_b(\mbf{q}_b,\mbf{u}_b) = \{\mbf{x}\in\mbb{R}^3\mid \mathbf{n}(\mbf{u}_b)^{\top}(\mathbf{x}-\mathbf{q}_{b})\leq  0\},
\end{align}
where the  normal vector of the $b$-th 6DMA surface $\mbf{n}(\mbf{u}_b)$ is given by
\begin{align}
	\mbf{n}(\mbf{u}_b) = \mbf{R}(\mbf{u}_b)\bar{\mbf{n}}. \label{n_def}
\end{align} 
As shown in Fig.\,\ref{fig:single_surface},  $\bar{\mbf{n}}$ in  (\ref{n_def}) is the predefined normal vector of a 6DMA surface in its local CCS, which aligns with the direction of its antennas' main lobe. Next, we define the  $b$-th 6DMA surface's occupied  region in the global CCS as 
\begin{align}
	\mbf{\mc{V}}_b(\mbf{q}_b,\mbf{u}_b) = \{\mbf{q}_b + \mbf{R}(\mbf{u}_b) \mbf{x}\mid \mbf{x}\in\mc{\bar{V}}\},
\end{align}
where $\mc{\bar{V}}$ (shown in Fig.\,\ref{fig:single_surface}) denotes the predefined 6DMA surface  region in its local CCS. Finally, we confine the placement of each 6DMA surface  within the non-positive halfspaces associated with all the other 6DMA surfaces, with the corresponding constraint  given by
\begin{align}
	\mbf{\mc{V}}_{{b'}}(\mbf{q}_{{b'}},\mbf{u}_{{b'}}) \subseteq \mc{H}^-_{b}(\mbf{q}_{b},\mbf{u}_{b}),~\forall {b} ,{{b'}} \in \mathcal{B}, b\neq b'. \label{P_feasible_c11}
\end{align}

\begin{figure}[!t]  
	\centering
	\includegraphics[width=2.1in]{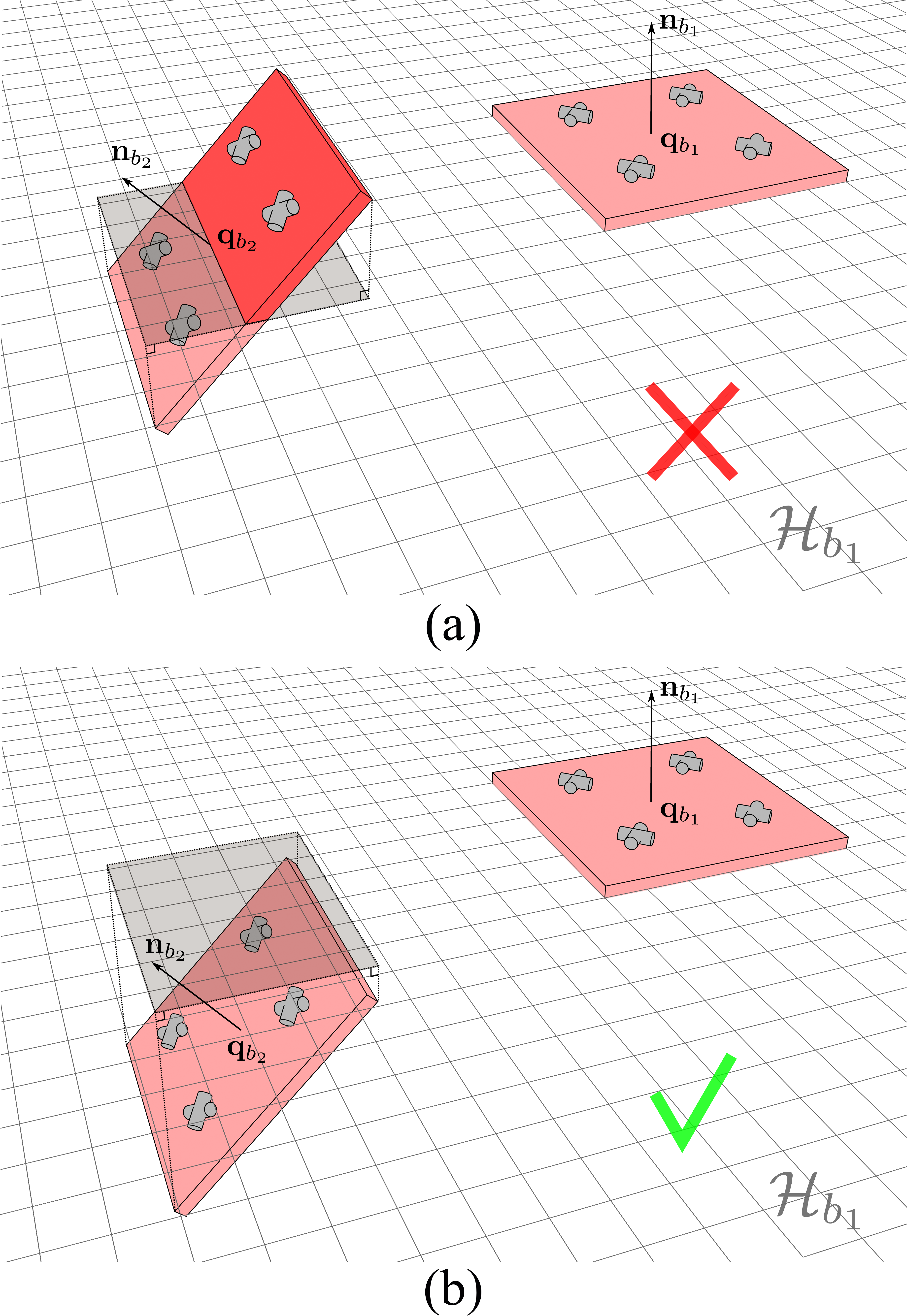}  
	\caption{Illustration of 6DMA surfaces' placement constraint (\ref{P_feasible_c11}). The boundary of the $b_1$-th surface's non-positive halfspace, $\mc{H}_{b_1}$, is indicated by the gray grid. In (a), the red-highlighted portion of the ${b_2}$-th surface region violates (\ref{P_feasible_c11}), as it blocks the $b_1$-th surface. In contrast,  (b) shows a valid configuration where  both the $b_1$-th and  ${b_2}$-th 6DMA surfaces are entirely within each other's non-positive halfspace, thereby preventing mutual signal blockage and overlap.}  
	\label{fig:constraint}  
\end{figure}

As illustrated in Fig.\,\ref{fig:constraint}, the constraint (\ref{P_feasible_c11}) effectively avoids the mutual signal blockage among different 6DMA surfaces since it prevents one  6DMA surface from being positioned in the front  halfspace of the others. 
Notably, the constraint (\ref{P_feasible_c11})  also avoids the overlap between any two 6DMA surfaces. This can be shown by contradiction. Suppose that two 6DMA  surfaces, indexed by $b$ and ${b'}$, overlap. Then, there must exist some points within the $b$-th surface's region that lie outside the ${b'}$-th surface's non-positive halfspace, and vice versa. This contradicts the constraint (\ref{P_feasible_c11}), thus ensuring that no overlap occurs.

\subsection{Problem Formulation}
The flexible adjustment of the 6DMA rotations and positions at  6DMA-BS can significantly enhance the communication performance of the users. In this paper, we aim to optimize the 6DMA position-rotation state $\mbf{z}$, for maximizing the   average sum log-rate of the multi-user (uplink) MAC by taking into account the rate fairness among the users. Assuming that the minimum mean-square error (MMSE) based linear receiver is applied at the 6DMA-BS to detect the users' signals independently, the  achievable average  rate of the $k$-th user can be written as
\begin{align}
	r_k(\mbf{\Sigma}(\mbf{z})) = \expa{\log_2 (1 + \mbf{h}_k^H \mbf{B}_k^{-1}\mbf{h}_k )}, \label{r_origin}
\end{align}
where the interference-plus-noise covariance matrix $\mbf{B}_k \in \mbb{C}^{BN \times BN}$ is given by
\begin{align}
	\mbf{B}_k = \sum_{\substack{k' \ne k \\ k' \in \mathcal{K}}} \frac{p_{k'}}{p_k}\mbf{h}_{k'}\mbf{h}_{k'}^H + \frac{\sigma^2}{p_k} \mbf{I}_{BN},\label{B_def}
\end{align}
where $p_1,\cdots,p_K$ represent the transmit power values of the users, respectively, and $\sigma^2$ denotes the average noise power at the BS's receiver. Note that in (\ref{r_origin}), the expectation is taken over all the users' random channels, $\mbf{h}_k$'s, given their SCI $\mbf{\Sigma}(\mbf{z})$. Accordingly, the objective function of our considered optimization problem is written as 
\begin{align}
	r(\mbf{z}) = \sum_{k=1}^K \log (r_k(\mbf{\Sigma}(\mbf{z}))). \label{obj_origin}
\end{align}

 However, it is hard to further derive the achievable  average rate in (\ref{r_origin}) due to the expectation over the nonlinear log function. A simple alternative  is to approximate this  expectation  by the Monte Carlo method\cite{6DMA_TWC}, i.e.,
\begin{align}
r_k(\mbf{\Sigma}(\mbf{z})) \approx \frac{1}{W}\sum_{w=1}^W {\log_2 \left(1 + {\mbf{h}_k^{w}}^H ({\mbf{B}_k^{w}})^{-1}\mbf{h}_k^{w} \right)}, \label{r_MC}
\end{align}
where  $\mbf{B}_k^{w}$  and  $\mbf{h}_k^{w}$ denote the $w$-th independently sampled values of the interference-plus-noise covariance matrix and channel of the $k$-th user, respectively, from the joint distribution of all the  users' channels  given by (\ref{hk_distri}), and $W$ is the number of Monte Carlo realizations with independently sampled channels. To provide an accurate approximation in (\ref{r_MC}), $W$ needs to be sufficiently large, which increases significantly with $K$ and $BN$, incurring higher  computational complexity for solving our formulated optimization problem. Therefore, in order to derive a more tractable expression as well as reduce the computational complexity, we first apply Jensen's inequality to  upper bound the  average rate by
\begin{align}
\bar{r}_k(\mbf{\Sigma}(\mbf{z}))  &= \log_2 \left(1 + \expa{\mbf{h}_k^H \mbf{B}_k^{-1}\mbf{h}_k} \right)  \nonumber \\
&= \log_2 \left(1 + \expa{\trace{\mbf{h}_k^H \mbf{B}_k^{-1}\mbf{h}_k}} \right)  \nonumber \\
&\overset{(a)}{=} \log_2 \left(1 + \trace{\expa{ \mbf{B}_k^{-1}\mbf{h}_k\mbf{h}_k^H}} \right) \nonumber \\
&\overset{(b)}{=} \log_2  \left( 1 + \trace{\expa{\mbf{B}_k^{-1}   } \mbf{\Sigma}_k(\mbf{z})   }  \right), \label{r_ub}
\end{align}
where $(a)$ is due to the linearity of trace and the property $\trace{\mbf{M}_1\mbf{M}_2} = \trace{\mbf{M}_2\mbf{M}_1}$ of trace, and $(b)$ follows from the definition $\mbf{\Sigma}_k(\mbf{z}) \triangleq \expa{\mbf{h}_k\mbf{h}_k^H}$ and the fact that $\expa{\mbf{M}_1\mbf{M}_2} = \expa{\mbf{M}_1}\expa{\mbf{M}_2}$ if $\mbf{M}_1$, $\mbf{M}_2$ are independent matrices  (which holds as $\mbf{h}_k$ is independent of $\mbf{h}_{k'}$'s for $k'\neq k$).
Note that computing $\expa{\mbf{B}_k^{-1}   }$ in (\ref{r_ub}) is still a non-trivial task, thus we again apply Jensen's inequality to lower bound (\ref{r_ub}) by
\begin{align}
 \underline{r}_k(\mbf{\Sigma}(\mbf{z})) = \log_2 \left( 1 + \trace{\expa{\mbf{B}_k }^{-1}   \mbf{\Sigma}_k(\mbf{z})  }  \right) , \label{r_ub_lb}
\end{align}
where  $\expa{\mbf{B}_k}$ can be easily derived as
\begin{align}
	\expa{\mbf{B}_k} =  \sum_{\substack{k' \ne k \\ k' \in \mathcal{K}}} \frac{p_{k'}}{p_k}\mbf{\Sigma}_{k'}(\mbf{z})  + \frac{\sigma^2}{p_k} \mbf{I}_{BN}.
\end{align}
By replacing the  average rate $r_k$ in (\ref{obj_origin}) by the approximation $\underline{r}_k$, the optimization problem is finally  formulated as
\begin{subequations}
\label{Sta_Opt}
\begin{align}
\text{(P1)}~~&~\mathop{\max}\limits_{\mathbf{\mbf{z}}}~~
 \sum_{k=1}^{K} \log(\underline{r}_k(\mbf{\Sigma}(\mbf{z}))) \label{P1_obj}\\
~&~	\mbf{\mc{V}}_{{b'}}(\mbf{q}_{{b'}},\mbf{u}_{{b'}}) \subseteq \mc{H}^-_{b}(\mbf{q}_{b},\mbf{u}_{b}),~\forall {b} ,{{b'}} \in \mathcal{B}, b\neq b', \label{P1c1}\\
~&~ \mathbf{q}_{b} \in \mathcal{V}_{\text{6DMA}}, ~\forall b  \in \mathcal{B}. \label{P1c3}
\end{align}
\end{subequations}

Notably,  $\underline{r}_k$ is a function of the SCI $\mbf{\Sigma}(\mbf{z})$\footnote{Assume that the users' transmit powers  and the noise power at the BS's receiver are known. }, which is determined by the position-rotation state  $\mbf{z}$ of the 6DMA-BS and the MPC information of all the  users. If the MPC information is known, the SCI $\mbf{\Sigma}(\mbf{z})$ can be obtained for arbitrary position-rotation state $\mbf{z}$ within the 6DMA-BS region, and so for $\underline{r}_k$'s. Therefore, in this paper we first estimate the MPC information of all  the  users to obtain their SCI in Section IV  and then solve the optimization problem (P1) over $\mbf{z}$ in Section V. Prior to them, we  propose  a practical  protocol for the 6DMA-BS to operate in the next section. 
\begin{figure*}[!t]  
	\centering
	\includegraphics[width=5.1in]{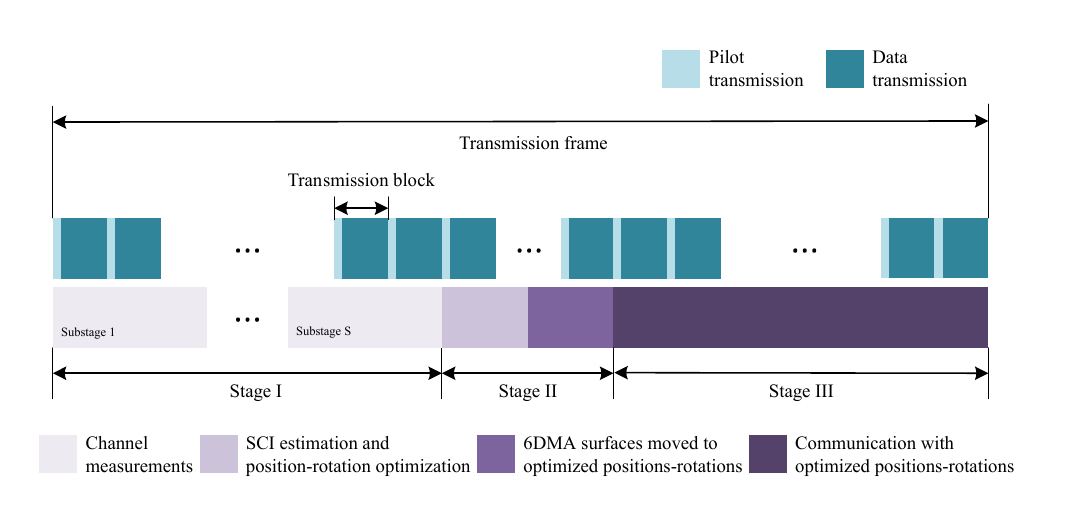}  
	\caption{The proposed three-stage protocol for  6DMA-BS.}  
	\label{fig:protocol}  
\end{figure*}
\section{6DMA-BS Protocol}
We propose a three-stage protocol  for the 6DMA-BS in this section. As shown in Fig.\,\ref{fig:protocol}, within a long transmission frame during which the SCI of users remains constant, the 6DMA-BS operates based on the following three stages.
\begin{itemize}
\item In Stage~I, the 6DMA-BS  takes channel measurements at   $M$ ($M\ge B$) training position-rotation pairs denoted by $\bar{\mbf{z}}_m = [\bar{\mbf{q}}_m^\top,\bar{\mbf{u}}_m^\top]^\top \in \mbb{R}^6$, $m\in\mc{M} \triangleq\{1,\cdots,M\}$, where $\bar{\mbf{q}}_m$ and $\bar{\mbf{u}}_m$ represent the $m$-th training position and rotation, respectively.
To perform the channel measurements at the $M$ training position-rotation pairs,  the $B$ 6DMA surfaces need to  change their position-rotation state, $\mbf{z}$, at least  $S = M/B$ times\footnote{Assume that $M$ is divisible by $B$ for convenience.}, with $S$ denoting the number of the substages shown in Fig.\,\ref{fig:protocol}. 
 Note that the moving speed of the 6DMA surfaces is much lower than that of the channel variation  in practice, thus each substage spans over a large number of transmission blocks, where the channel is assumed to be constant during each transmission block.
 As shown in Fig.\,\ref{fig:Trainings}, at the $s$-th substage, $s \in \mc{S} \triangleq\{1,\cdots,S\}$, the $B$ 6DMA surfaces update their position-rotation state   to  $\mbf{z}^{(s)} \triangleq  [\bar{\mbf{z}}_{B(s-1)+1},\cdots,\bar{\mbf{z}}_{Bs}] \in \mbb{R}^{6B}$, estimate and store the instantaneous channels of all the transmission blocks within the  substage based on the received pilot signals sent by all the users. 
 \begin{figure}[!t]  
 	\centering
 	\includegraphics[width=2.5in]{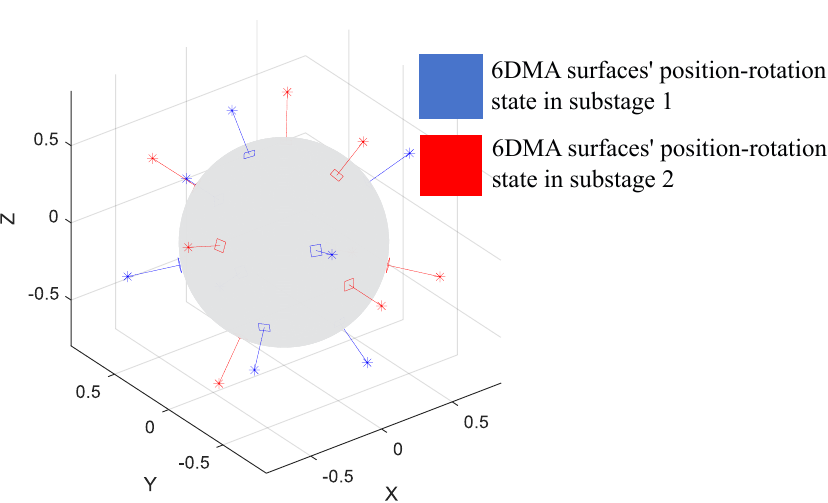}  
 	\caption{Illustration of 6DMA surfaces' position-rotation state in different substages of Stage I. Assume $B=8$ and $M=2B=16$ training position-rotation pairs for channel measurement, which are generated according to (\ref{Fibonacci_q}) and (\ref{Fibonacci_u}). Since $S=M/B = 2$, 6DMA surfaces are  moved to the position-rotation state indicated in blue in the first substage  to estimate $\mbf{{h}}_k^{(1)}$ and then to  that  indicated in red in the second substage to estimate $\mbf{{h}}_k^{(2)}$.}  
 	\label{fig:Trainings}  
 \end{figure}
\item In Stage~II, the 6DMA-BS estimates the MPC information of all the users based on the channel measurements taken in Stage~I. Then, based on the MPC information, the 6DMA-BS reconstructs all the users' SCI, $\mbf{\Sigma}(\mbf{z})$, for any position-rotation state $\mbf{z}$  within the 6DMA-BS region and  solves the optimization problem (P1) accordingly.
Subsequently, the 6DMA-BS moves its 6DMA surfaces to their optimized position-rotation state, denoted by $\mbf{z}^*$. 

\item In Stage~III, the 6DMA-BS communicates with the users given the optimized position-rotation state, $\mbf{z}^*$, for the remainder of the transmission frame.
 \end{itemize}

It is worth noting that the proposed protocol is easily implementable in existing wireless networks as it does not require modifications to the existing communication protocol of BSs over channel coherence (transmission)  blocks. From the users' perspective, their communications with the 6DMA-BS are conducted normally during the whole transmission frame, without any interruption. Nevertheless, there are slowly-varying channel conditions for users due to the 6DMA surfaces' (slow) movement in Stages I and II, while they will experience average rate improvement in Stage~III after the 6DMA surfaces are moved to their optimized position-rotation state. 
\section{Estimation of SCI}
In this section, we propose two methods to design the training  position-rotation pairs for the channel measurements in Stage~I and   estimate  the MPC information of all the users in Stage~II, respectively.  
Then, we discuss the minimum  value of $M$ needed to accurately resolve the MPCs of all the users.
\subsection{Training Position-Rotation Pair Design}
To accommodate all possible  DoAs of the users, the training position-rotation pairs for the channel measurements in Stage~I of the proposed protocol should be uniformly distributed. Specifically, we apply the Fibonacci method  to generate the training positions\cite{6DMA_TWC}. First, we express the training positions, $\bar{\mbf{q}}_m$, $m \in\mc{M}$, in the spherical coordinate system as $\bar{\mbf{q}}_m = [\sin\theta_m \cos \phi_m, \sin\theta_m \sin \phi_m,\cos\theta_m ]$. Then, the spherical coordinates of $\bar{\mbf{q}}_m$ are given by
\begin{align}
	\theta_m &= \arccos\left(1 - \frac{2(m + 0.5)}{M}\right),\\
	\phi_m &= m\frac{2\pi}{\phi_g} ~\bmod ~2\pi,\label{Fibonacci_q}
\end{align}
where $\phi_g = \frac{1+\sqrt{5}}{2}$ is the golden ratio, and $\bmod$ represents the modulo operation. For the training rotations, we select $\bar{\mbf{u}}_m$, $m \in\mc{M}$, such that 
\begin{align}
\mbf{n}(\bar{\mbf{u}}_m ) = \bar{\mbf{q}}_m, \label{Fibonacci_u}
\end{align}
where $\mbf{n}(\bar{\mbf{u}}_m)$ is the normal vector of a 6DMA surface with rotation $\bar{\mbf{u}}_m$. 
\subsection{MPC Information Estimation}
Next, we  introduce a reduced-sample (RS) method for the MPC estimation. Specifically, the RS method reuses the $B$  6DMA surfaces across the $S$ substages ($S\ge 1$) to perform  channel measurements in Stage~I of the proposed protocol, thereby enabling more spatial sampling than that achieved by a static deployment of $B$ 6DMA surfaces. Based on these channel measurements, an OMP-based algorithm is then employed to jointly estimate the users' DoAs, $\mbf{f}_{k,l}$'s,  and their respective average power values, $\mbf{D}_k$'s.

More specifically, for the $k$-th user, $k\in\mc{K}$, its instantaneous channel to the $B$  6DMA surfaces in the $t$-th transmission block of the $s$-th  substage,  $s \in \mc{S}$, is denoted  by $\mbf{{h}}^{(s)}_k[t]\in\mbb{C}^{BN}$, $t\in\mc{T} \triangleq \{1,\cdots,T\}$. Note that in the $s$-th substage, the $B$ 6DMA surfaces communicate with the users with the position-rotation state $\mbf{z}^{(s)}= [\bar{\mbf{z}}_{B(s-1)+1},\cdots,\bar{\mbf{z}}_{Bs}]$.  Following (\ref{R_def}), the  covariance matrix   of the channel from the $k$-th user to the $B$  6DMA surfaces in the $s$-th substage, $\mbf{\Sigma}_{\mbf{h}_k^{(s)}} \triangleq \expa{\mbf{h}_k^{(s)}(\mbf{h}_k^{(s)})^{H}}\in \mbb{C}^{BN \times BN}$,  can be written as
\begin{align}
	\mbf{\Sigma}_{\mbf{h}_k^{(s)}} &= \mbf{\Sigma}_k(\mbf{z}^{(s)}) \nonumber \\
	&= \tilde{\mbf{A}}(\mbf{z}^{(s)},\mbf{F}_k) \mbf{D}_k \tilde{\mbf{A}}(\mbf{z}^{(s)},\mbf{F}_k)^H. 
\end{align}
Meanwhile, since all the users' instantaneous channels,  $\mbf{{{h}}}_k^{(s)} [t]$'s,  are estimated from the received pilot signals sent by the users\cite{textbook}, the covariance matrix $\mbf{\Sigma}_{\mbf{h}_k^{(s)}}$ can be approximated by the time average as
\begin{align}
\mbf{\Sigma}_{\mbf{h}_k^{(s)}} &\approx \frac{1}{T}  \sum_{t=1}^T
 \mbf{{h}}_k^{(s)}[t] \left(\mbf{{h}}_k^{(s)}[t]\right)^H \triangleq \hat{\mbf{\Sigma}}_k^{(s)}. \label{R_h_bar_apporx}
\end{align} 
By considering all substages and ignoring the approximation error in (\ref{R_h_bar_apporx}), we obtain the following $S$ observations  from which the users' DoAs and their respective average power values can be estimated.
\begin{align}
\hat{\mbf{\Sigma}}_k^{(s)} = \mbf{\Sigma}_{\mbf{h}_k^{(s)}},~s\in\mc{S}. \label{origin_OMP_eqs}
\end{align}

Next, we rewrite the equations in (\ref{origin_OMP_eqs}) to match the form of compressed sensing (CS) problems. Note that $\mbf{D}_k = \text{diag}(\alpha_{k,1},\cdots, \alpha_{k,L_k})$. Thus, for each substage $s$, we have
\begin{align}
	\hat{\mbf{\Sigma}}_k^{(s)}  = \sum_{l=1}^{L_k} \alpha_{k,l} \tilde{\mbf{a}}(\mbf{z}^{(s)},\mbf{f}_{k,l}) \tilde{\mbf{a}}(\mbf{z}^{(s)},\mbf{f}_{k,l})^H, \label{R_k^s_matrix}
\end{align}
where the weighted steering vector $\tilde{\mbf{a}}(\mbf{z}^{(s)},\mbf{f}_{k,l})$ is given by (\ref{weighted_steering_vector}). By vectorizing the matrices on  both sides, (\ref{R_k^s_matrix}) can be equivalently written as
\begin{align}
	\text{vec}\left(\hat{\mbf{\Sigma}}_k^{(s)}\right)  &= \sum_{l=1}^{L_k} \alpha_{k,l}\text{vec}\left( \tilde{\mbf{a}}(\mbf{z}^{(s)},\mbf{f}_{k,l}) \tilde{\mbf{a}}(\mbf{z}^{(s)},\mbf{f}_{k,l})^H\right) \nonumber \\
	&= \mbf{A}_{\text{vec}}^{(s)}(\mbf{z}^{(s)},\mbf{F}_k)\bm{\alpha}_k,
\end{align}
where the  $l$-th column of $\mbf{A}_{\text{vec}}^{(s)}(\mbf{z}^{(s)},\mbf{F}_k) \in \mbb{C}^{B^2N^2\times L_k}$, $\forall l \in \mc{L}_k$, is defined as $\text{vec}\left( \tilde{\mbf{a}}(\mbf{z}^{(s)},\mbf{f}_{k,l}) \tilde{\mbf{a}}(\mbf{z}^{(s)},\mbf{f}_{k,l})^H\right)$, and $\bm{\alpha}_k \triangleq [\alpha_{k,1}, \cdots,\alpha_{k,L_k}]^\top\in \mbb{R}_+^{L_k}$. By combining all the $S$ equations, we have
\begin{align}
	\text{vec}\left(\hat{\mbf{\Sigma}}_k\right) = \mbf{A}_{\text{vec}} (\bar{\mbf{z}},\mbf{F}_k)\bm{\alpha}_k, \label{OMP_form}
\end{align}
where   $\text{vec}\left(\hat{\mbf{\Sigma}}_k\right) \in \mbb{C}^{B^2N^2S}$ is defined as the equivalent received signal vector by collecting the $k$-th user's vectorized covariance  matrices  in all the $S$ substages, with $\hat{\mbf{\Sigma}}_k  \triangleq  \left[\hat{\mbf{\Sigma}}_k^{(1)},\cdots,\hat{\mbf{\Sigma}}_k^{(S)}\right] \in \mbb{C}^{BN\times BNS}$; and $ \mbf{A}_{\text{vec}} (\bar{\mbf{z}},\mbf{F}_k)  \in \mbb{C}^{B^2N^2S \times L_k}$ is a dictionary matrix formed by $\mbf{A}_{\text{vec}}^{(s)}(\mbf{z}^{(s)},\mbf{F}_k)$'s, $\forall s$, i.e.,
\begin{align}
 \mbf{A}_{\text{vec}} (\bar{\mbf{z}},\mbf{F}_k) &= \left[\mbf{A}_{\text{vec}}^{(1)} (\mbf{z}^{(1)},\mbf{F}_k)^H,\cdots,\mbf{A}_{\text{vec}}^{(S)}(\mbf{z}^{(S)},\mbf{F}_k)^H \right]^H,
\end{align}
with $\bar{\mbf{z}} \triangleq [(\mbf{z}^{(1)})^\top,\cdots,(\mbf{z}^{(S)})^\top]^\top\in \mbb{R}^{6BS}$. Next, we discretize  possible values of DoA coordinates into a finite set. Let  $\mbf{F}^g \triangleq [\mbf{f}^g_{1},\cdots,\mbf{f}^g_{I}]\in \mbb{R}^{3\times I}$ denote a  discretization set, where $I$ denotes the  size of the discrete set from which the  DoAs take values, and  the columns of $\mbf{F}^g$ represent all the candidate coordinates of the DoAs. In particular,  the columns of $\mbf{F}^g$ can be chosen using the Fibonacci method proposed in Section~IV-A. By ignoring the quantization error, the columns of $\mbf{A}_{\text{vec}}(\bar{\mbf{z}},\mbf{F}_k) \in \mbb{C}^{B^2N^2S \times L_k}$ in (\ref{OMP_form}) can be viewed as a subset of the columns of $\mbf{A}_{\text{vec}}(\bar{\mbf{z}},\mbf{F}^g)  \in \mbb{C}^{B^2N^2S \times I}$, and
\begin{align}
	\mbf{A}_{\text{vec}} (\bar{\mbf{z}},\mbf{F}_k)\bm{\alpha}_k = \mbf{A}_{\text{vec}} (\bar{\mbf{z}},\mbf{F}^g)\bm{\alpha}^g_k,
\end{align}
where $\bm{\alpha}^g_k \in \mathbb{R}^I$ is a non-negative $L_k$-sparse vector, whose non-zero entries represent the average powers of the MPCs of user $k$, and whose support  indices correspond to the DoAs specified by the columns of $\mathbf{F}^g$. Accordingly,  we  formulate the following CS problem for the MPC estimation,
\begin{subequations}
	\begin{align}
		\text{(P2)}~&~ \min_{\bm{\alpha}_k^g\succeq 0} \left\| \mbf{A}_{\text{vec}} (\bar{\mbf{z}},\mbf{f}^g) \bm{\alpha}^g_k-	\text{vec}\left(\hat{\mbf{\Sigma}}_k\right)    \right\|^2_2\\
		\text{s.t.}~&~  \| \bm{\alpha}_k^g \|_0 = L_k,
	\end{align}
\end{subequations}
which can be efficiently solved by the non-negative OMP algorithm \cite{OMP}.
Let ${\bm{\alpha}_k^g}^*$ denote the solution to (P2), and let $\mathcal{I}$ be the support set of its nonzero entries. Then, the user's DoAs and their respective average power values are estimated as 
\begin{align}
	\mbf{F}_k^* &= [\mbf{F}^g]_{:,\mc{I}},\\
	\mbf{D}_k^* &=\text{diag}\left([{\bm{\alpha}_k^g}^*]_\mc{I}\right),
\end{align}
respectively, where $[\mbf{F}^g]_{:,\mathcal{I}}$ denotes the submatrix of $\mbf{F}^g$ formed by extracting all the rows and the columns indexed by the set $\mathcal{I}$, and $[{\bm{\alpha}_k^g}^*]_\mc{I}$ denotes the entries of ${\bm{\alpha}_k^g}^*$ indexed by $\mathcal{I}$.

\subsection{Minimum Number of Training Position-Rotation Pairs}
In this subsection, we analyze the lower bound on the number of training position-rotation pairs, $M_{\text{min}}$, required to accurately resolve the users' MPC information. In the  conventional OMP algorithm applied to antenna arrays with the isotropic (unit) antenna gain, the total number of antennas needs to be no less than the number of multiple paths to ensure an accurate and robust estimation performance,  i.e.,
\begin{align}
	M_{\text{min}}N =  C \max_k\{L_k\}, \label{uniform_MN}
\end{align}
where $C\ge 1$ is a coefficient related to the OMP algorithm \cite{OMP}.
For DoA estimation in the case of 6DMA-BS, however, the directional radiation pattern of its antennas may render some of the 6DMA surfaces ineffective to measure the channels from the users if their path DoAs are outside the  6DMA surfaces' coverage regions where the antenna gain is practically significant.    To ensure that at least $C\max_k\{L_k\}$ antennas of the 6DMA-BS are effective for arbitrary signal direction,  more training position-rotation pairs should be employed. Specifically, we require that
\begin{align}
\left\lfloor\frac{M_{\text{min}}}{\beta(\Theta)}\right\rfloor N = C \max_k\{L_k\}, \label{M_require}
\end{align}
where $\Theta\in \left(0,\pi\right]$ denotes the beamwidth of directional antennas on the 6DMA surfaces, and $\beta(\Theta)$ denotes the minimum number of cones with apex angle $\Theta $ required to fully cover a sphere, provided that the cones and the sphere are centered at the same point. Note that (\ref{M_require}) assumes an ideal scenario where the orientations of the training pairs are perfectly aligned with the optimal cone placement that minimizes \( \beta(\Theta) \). In general, such alignment may not be achievable. Therefore, we derive a more practical lower bound on the required number of 6DMA training position-rotation pairs as
\begin{align}
	M_{\text{min}} \ge \left\lceil \frac{C \beta(\Theta) \max_k\{L_k\}}{N} \right\rceil. \label{M_bound}
\end{align}
This expression indicates that narrower beams require a larger \( M \) to sufficiently cover the spatial domain and ensure accurate resolution of MPC information  (as will be shown in Section VI by simulation).

\section{6DMA Position and Rotation Optimization Based on SCI }
In this section, we propose an efficient algorithm to solve (P1), based on the estimated SCI of users at the 6DMA-BS, as in Stage~II of the proposed protocol. The main challenge for solving (P1) lies in the closely coupled position and rotation variables for the 6DMA surfaces in its objective function, which makes the problem difficult to be solved optimally. To tackle this challenge, we first relax (P1) to an unconstrained optimization problem to optimize  6DMA surfaces' rotations, $\mbf{u}_b$'s, only by expressing their positions, $\mbf{q}_b$'s, as a function of the corresponding rotation of each 6DMA surface. Then,  with the optimized rotations, we proceed to find the feasible positions of 6DMA surfaces that can realize such rotations under practical placement constraints. As compared to the existing alternating optimization based  algorithm  in \cite{6DMA_TWC}, which optimizes the positions and rotations of 6DMA surfaces alternately in an iterative manner, our proposed sequential optimization algorithm decouples the rotation and position optimizations, thus it does not require alternating optimization with iterations and achieves much lower computational complexity. Moreover, it avoids the common issue with alternating optimization  on getting trapped in undesired local optima, thereby improving the performance of the solution obtained. Notably, the superior performance of the proposed algorithm also reveals that the rotation adjustment of 6DMA surfaces in fact contributes more significantly to the performance enhancement over their position adjustment, providing key insight into the performance enhancement mechanism of 6DMA systems over conventional FPAs.     

\subsection{Rotation Optimization}
In this subsection, we relax (P1) to an optimization problem over  the rotations of the 6DMA surfaces only.  First, we drop the unified blockage and overlap constraint (\ref{P1c1}). Note that the remaining constraint ensures that the feasible surfaces lie within the 6DMA region, while a spherical region provides a simple construction that satisfies this condition. To  confine the 6DMA surfaces to a spherical region, we set the position-rotation pair of each 6DMA surface as
\begin{align}
	\mbf{z}_b(\mbf{u}_b) = [d_{\text{ins}} \mbf{n}(\mbf{u}_b)^\top, \mbf{u}_b^\top]^\top, ~\forall b \in \mc{B}, \label{zb_over_ub}
\end{align}
where $d_{\text{ins}}$ denotes the largest possible radius of an inscribed sphere inside the 6DMA region, $\mc{V}_{\text{6DMA}}$. Accordingly, the  position-rotation state of the $B$ 6DMA surfaces takes the form
\begin{align}
	\mbf{z}(\mbf{u}) =[\mbf{z}_1(\mbf{u}_1),\cdots,\mbf{z}_B(\mbf{u}_B)], \label{z_over_u}
\end{align}
where $\mbf{u} \triangleq [\mbf{u}_1^\top,\cdots,\mbf{u}_B^\top]^\top\in\mbb{R}^{3B}$.
 Substituting (\ref{z_over_u}) into (\ref{P1_obj}), (P1) is relaxed  into the following unconstrained optimization problem:
\begin{subequations}
	\label{P3A}
	\begin{align}
		\text{(P3-A)}~~&~\mathop{\max}\limits_{\mathbf{u}}~~
		\sum_{k=1}^{K} \log(\underline{r}_k(\mbf{\Sigma}( \mbf{z}(\mbf{u})))).
	\end{align}
\end{subequations}
\begin{remark}
	The rationale for formulating  (P3-A) lies in the fact  that the rotations of 6DMA surfaces  are the dominant factor that affects the channel covariance matrix (SCI),  $\mbf{\Sigma}_k(\mbf{z})$'s,  of the users,  shown as follows. Denote  the $(i,j)$-th entry of $\mbf{\Sigma}_k(\mbf{z})$ and the $(i,j)$-th entry of $\tilde{\mbf{A}}(\mbf{z},\mbf{F}_k)$ by  $r_{i,j}^k$ and $a_{i,j}^k$ respectively. From (\ref{hkb}), we can derive
	\begin{align}
		r_{i,j}^k
		&= \sum_{l=1}^{L_k} \alpha_{k,l}^2 a_{i,l}^k{a_{j,l}^{k \dag}} \nonumber\\
		&= \sum_{l=1}^{L_k} \alpha_{k,l}^2 \sqrt{g(\mbf{u}_{\lfloor\frac{i}{N}\rfloor},\mbf{f}_{k,l})g(\mbf{u}_{\lfloor\frac{j}{N}\rfloor},\mbf{f}_{k,l})} e^{j(\angle{a_{i,l}^k} - \angle{a_{j,l}^k})},
	\end{align}
	where $\angle(\cdot)$ denotes the phase of a complex number. It is observed that antenna rotations influence the amplitude of the channel covariance matrix  via the antenna gain. In a typical 6DMA system equipped with directional antennas, the impact of antenna gain is significant, thus optimizing the rotations of the 6DMA surfaces is crucial to the achievable rate of 6DMA systems.
\end{remark}

 Next, we apply a gradient descent algorithm to solve (P3-A). Denote the objective function in (P3-A) by $\tilde{r}(\mbf{u})$. The gradient of $\tilde{r}(\mbf{u})$ can be calculated numerically based on its definition. Specifically, define $\mbf{e}_i \in \mbb{R}^{3B}$ as a vector whose $i$-th entry is 1 and otherwise 0, and $\epsilon$ as a small positive number. Then, the partial derivative of $\tilde{r}(\mathbf{u})$ w.r.t. the $i$-th entry of $\mbf{u}$, denoted by $u_i$, $i=1,\cdots,3B$, is approximated by
\begin{align}
	\frac{\partial \tilde{r}(\mbf{u})}{\partial u_i} \approx \left(\tilde{r}(\mbf{u} + \epsilon   \mbf{e}_i) -\tilde{r}(\mbf{u}) \right)/\epsilon, \label{Nabla_1}
\end{align}
and the corresponding gradient is given by
\begin{align}
	\frac{\partial \tilde{r}(\mbf{u})}{\partial \mbf{u}} \triangleq [\frac{\partial \tilde{r}(\mbf{u})}{\partial {u}_1},\cdots,\frac{\partial \tilde{r}(\mbf{u})}{\partial {u}_{3B}}]^\top \label{Nabla_2}.
\end{align}
In each iteration, the descent direction of the rotations of the 6DMA surfaces is set to $\frac{\partial \tilde{r}(\mbf{u})}{\partial \mbf{u}}$, and the corresponding step size is determined by the Armijo rule \cite{armijo}. The proposed algorithm starts with an initial vector  $\mbf{u}^{(0)}$ (to be specified later) and generates a sequence of vectors $\{\mbf{u}^{(\kappa)}\}$, $\kappa=1,\cdots,\kappa_{\text{max}}$, as
\begin{align}
\mbf{u}^{(\kappa)} = \mbf{u}^{(\kappa-1)} + \tau^{(\kappa-1)}\left. \frac{\partial \tilde{r}(\mbf{u})}{\partial \mbf{u}}\right|_{\mbf{u}^{(\kappa-1)}}.
\end{align}
The step size $\tau^{(\kappa-1)}$ in the $(\kappa-1)$-th iteration is calculated as $\tau_{\text{ini}} \delta^\nu$, where $\tau_{\text{ini}}$ denotes the initial step size, $\delta \in (0,1)$, and $\nu$ denotes the smallest nonnegative integer which satisfies
\begin{align}
\tilde{r}(\mbf{u}^{(\kappa)}) - &\tilde{r}(\mbf{u}^{(\kappa-1)}) > \nonumber\\ 
& \tau_{\text{ini}} \delta^\nu  \left. \frac{\partial \tilde{r}(\mbf{u})}{\partial \mbf{u}}\right|_{\mbf{u}^{(\kappa-1)}}^\top (\mbf{u}^{(\kappa)} - \mbf{u}^{(\kappa-1)}). \label{Armijo}
\end{align}

For initialization, we propose a greedy search algorithm to select the initial rotation vector, $\mbf{u}^{(0)}$, of the 6DMA surfaces to ensure good performance of the converged solution. First, let $\mc{C} \subset \mbb{R}^3$ denote a candidate set consisting of $\bar{M}$ rotations uniformly generated using the Fibonacci method. The initial rotation vector is then selected from $\mc{C}$ in an iterative manner. Specifically, denote the rotations selected up to the $(\kappa'-1)$-th iteration, where $\kappa' = 1, \cdots, B$, by $\hat{\mbf{u}}^{(\kappa'-1)} \in \mbb{R}^{3(\kappa'-1)}$. In the $\kappa'$-th iteration, we perform an exhaustive search to identify the optimal $\hat{\mbf{u}}^* \in \mbb{R}^3$ from the candidate set $\mc{C}$ that maximizes the objective function of (P3-A), i.e.,
\begin{align}
	\hat{\mbf{u}}^* \leftarrow \arg\max_{\underline{\mbf{u}} \in \mc{C}} \tilde{r}([(\hat{\mbf{u}}^{(\kappa'-1)})^\top, \underline{\mbf{u}}^\top]^\top). \label{greedy_search_adaptation}
\end{align}
This procedure is repeated until all the rotations are selected for the $B$ 6DMA surfaces.

\subsection{Position Optimization}
\begin{figure}[!t]  
	\centering
	\includegraphics[width=2.5in]{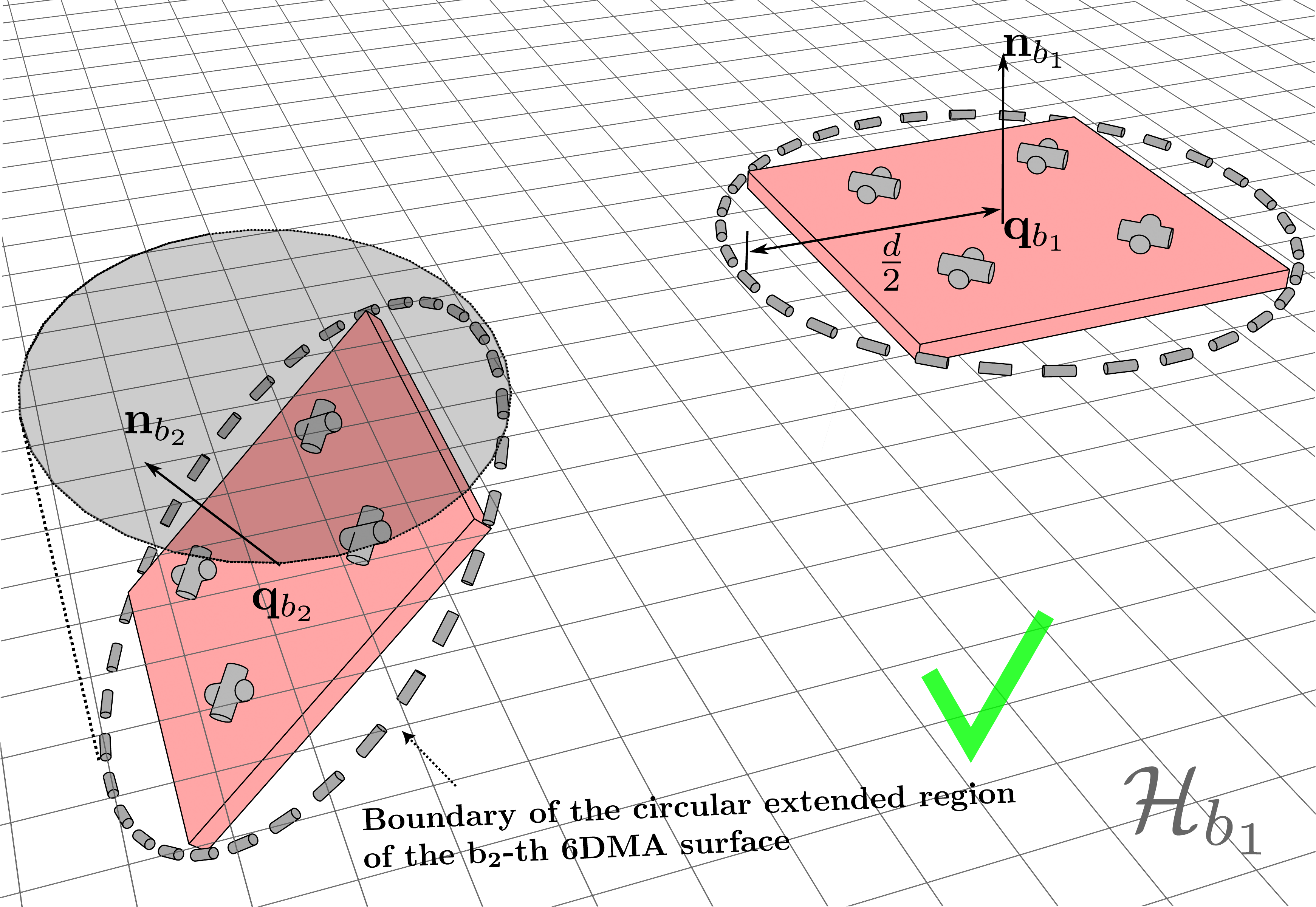}  
	\caption{Illustration of the tightened constraint (\ref{P_feasible_c1_new}). The boundary of the non-positive hyperplane associated with the $b_1$-th 6DMA surface, $\mc{H}_{b_1}$, is indicated by the gray grid.  The $b_1$-th and ${b_2}$-th 6DMA surfaces satisfy the tightened constraint (\ref{P_feasible_c1_new}), as both of  their CERs  are entirely within the non-positive halfspace associated with  each other.}  
	\label{fig:extended_constraint}  
\end{figure}

Next, we aim to find feasible  positions of the 6DMA surfaces subject to their optimized rotations obtained in Section V-A. Denote the optimized rotations of 6DMA surfaces after solving (P3-A) by $\mbf{u}^* = [\mbf{u}_1^{*\top},\cdots,\mbf{u}_B^{*\top}]^\top \in \mbb{R}^{3B}$. In order to make the optimized rotations $\mbf{u}^*$ practically implementable, we formulate the following feasibility problem, i.e., 
\begin{subequations}
	\label{feasible}
	\begin{align}
		\text{(P3-B)}~&~\mathop{\text{Find}}~~\mbf{q}\in\mbb{R}^{3B}~~\text{such that}\nonumber\\
		&~\mc{V}_{{b'}}(\mbf{q}_{{b'}},\mbf{u}_{{b'}}^*) \subseteq \mc{H}_{b}^-(\mbf{q}_{b},\mbf{u}_{b}^*) ,~\forall {b} ,{{b'}} \in \mathcal{B}, b\neq b', \label{P_feasible_c1}\\
		&~ \mathbf{q}_{b} \in \mathcal{V}_{\text{6DMA}}, ~\forall b  \in \mathcal{B}, \label{P_feasible_c3}
	\end{align}
\end{subequations}
where $\mbf{q} \triangleq [\mbf{q}_1^\top,\cdots,\mbf{q}_B^\top]^\top\in\mbb{R}^{3B}$. Notably, the constraint (\ref{P_feasible_c1}) is highly non-convex, as the derivation of each position in $\mc{V}_b(\mbf{q}_b,\mbf{u}_b^*)$ involves the multiplication with the rotation matrix $\mbf{R}(\mbf{u}_b^*)$ whose elements are  sinusoidal polynomials over $\mbf{u}_b^*$.
To cope with this non-convex optimization problem, we propose a geometry-based algorithm to search for the feasible positions of the $B$ 6DMA surfaces (with their rotations given) by iteration. 
First, we tighten  constraint (\ref{P_feasible_c1}) to simplify the positioning algorithm. 
Specifically, we replace the constraint (\ref{P_feasible_c1}) by
\begin{align}
	\mc{V}^{\text{e}}_{{b'}}(\mbf{q}_{{b'}}) \subseteq \mc{H}_{b}^-(\mbf{q}_{b},\mbf{u}_{b}) ,~\forall {b} ,{{b'}} \in \mathcal{B}, b\neq b', \label{P_feasible_c1_new}
\end{align}
where the circular extended region (CER) of the $b$-th 6DMA surface, $\mc{V}^{\text{e}}_b(\mbf{q}_b)\subset\mbb{R}^3$, is defined as a circle centered at $\mbf{q}_b$, with normal vector $\mbf{n}^*_b$ and radius $d/2$, i.e.,
\begin{align}
	\mc{V}^{\text{e}}_b(\mbf{q}_b) \triangleq \{\mbf{x}\in\mbb{R}^3\mid \|(\mbf{I}_3 - \mbf{n}^*_b \mbf{n}^{*\top}_b)(\mbf{x}-\mbf{q}_b)\|_2 \le \frac{d}{2}\} \label{CER_def}.
\end{align}
Note that the radius of the CER,  \( \frac{d}{2} \), is chosen as the minimum radius required to enclose the 6DMA surface region. For example, for square surfaces with edge length \( d' \), the CER radius is \( \frac{\sqrt{2}}{2}d' \). As shown in Fig.\,\ref{fig:extended_constraint}, since  $\mbf{\mc{V}}_b(\mbf{q}_b,\mbf{u}^*_b) \subseteq \mc{V}^{\text{e}}_b(\mbf{q}_b)$,   (\ref{P_feasible_c1_new})  guarantees that a larger region of each 6DMA surface lies within in the non-positive halfspace associated with all the other surfaces. 

Next, we define $\mc{X}^{(\kappa)}$ and $\mc{Y}^{(\kappa)}$ as the index sets which include the indices of the already positioned  6DMA surfaces and the remaining non-positioned  6DMA surfaces at the $\kappa$-th iteration, $\kappa=1,\cdots,B$, respectively. Thus, $\mc{X}^{(0)} $ is an empty set and $\mc{Y}^{(0)} = \mc{B}=\{1,\cdots,B\}$.  
At the first iteration, we randomly select a surface and assign its position arbitrarily, for example, the coordinate origin. At the $\kappa$-th iteration, we position the  surface whose normal vector is most closely aligned with those  in $\mc{X}^{(\kappa-1)}$. Specifically, the index of the surface to be positioned is given by
\begin{align}
	b^{(\kappa)} = \arg\mathop{\max}_{b\in \mc{Y}^{(\kappa-1)}}  \{ \max_{{b'}\in \mc{X}^{(\kappa-1)}} {\mbf{n}^*_{b}}^\top \mbf{n}^*_{{b'}} \}.
\end{align}
Then, we jointly position the $b^{(\kappa)}$-th 6DMA surface along with the surfaces indexed by $\mc{X}^{(\kappa-1)}$, while ensuring that  the constraint ($\ref{P_feasible_c1_new})$ is satisfied. The positioning policy for the $b^{(\kappa)}$-th 6DMA surface consists of three steps in each iteration. Specifically, the operations of the $\kappa$-th iteration include: 
\begin{itemize}[]
\item{Step 1:} Select the hyperplane (denoted by $\mc{H}^{(\kappa)}$) on which the $b^{(\kappa)}$-th surface is to be placed. Specifically, the normal vector of $\mc{H}^{(\kappa)}$ coincides with that of the  $b^{(\kappa)}$-th 6DMA surface. Moreover, the CERs associated with 6DMA surfaces  in $\mc{X}^{(\kappa-1)}$ should lie within the non-positive halfspace of  $\mc{H}^{(\kappa)}$,  and  the  distance between these CERs and $\mc{H}^{(\kappa)}$ should be minimized. Thus, the hyperplane $\mc{H}^{(\kappa)}$ is given by
  \begin{align}
  	\mc{H}^{(\kappa)} = \left\{\mbf{x}\in\mbb{R}^3\mid {\mbf{n}^{*\top }_{b^{(\kappa)}} } \mbf{x} = \chi^{(\kappa)} \right\}, \label{get_plane1}
  \end{align}
  where $\chi^{(\kappa)}$ is given by
\begin{align}
\chi^{(\kappa)}  &= \max_{b \in \mc{X}^{(\kappa-1)}} \max_{\mbf{x}\in \mc{V}^{\text{e}}_{b}(\mbf{q}_{b})}  {\mbf{n}^{*\top}_{b^{(\kappa)}} } \mbf{x} . \label{Compute_r}
\end{align} 
Equation (\ref{Compute_r}) ensures that at least one of the CERs associated with 6DMA surfaces  in $\mc{X}^{(\kappa-1)}$ is tangent to the hyperplane $\mc{H}^{(\kappa)}$. We denote the index of the  tangent CER and the position of its corresponding tangent point as
\begin{align}
		b^{(\kappa)}_{\text{tan}},\mbf{x}_{\text{tan}}^{(\kappa)}&= \arg\mathop{\max}_{b \in \mc{X}^{(\kappa-1)}, \mbf{x}\in \mc{V}^{\text{e}}_{b}(\mbf{q}_{b})}  {\mbf{n}^{*\top}_{b^{(\kappa)}} } \mbf{x}.
\end{align} 

\begin{figure}[!t]  
	\centering
	\includegraphics[width=2.5in]{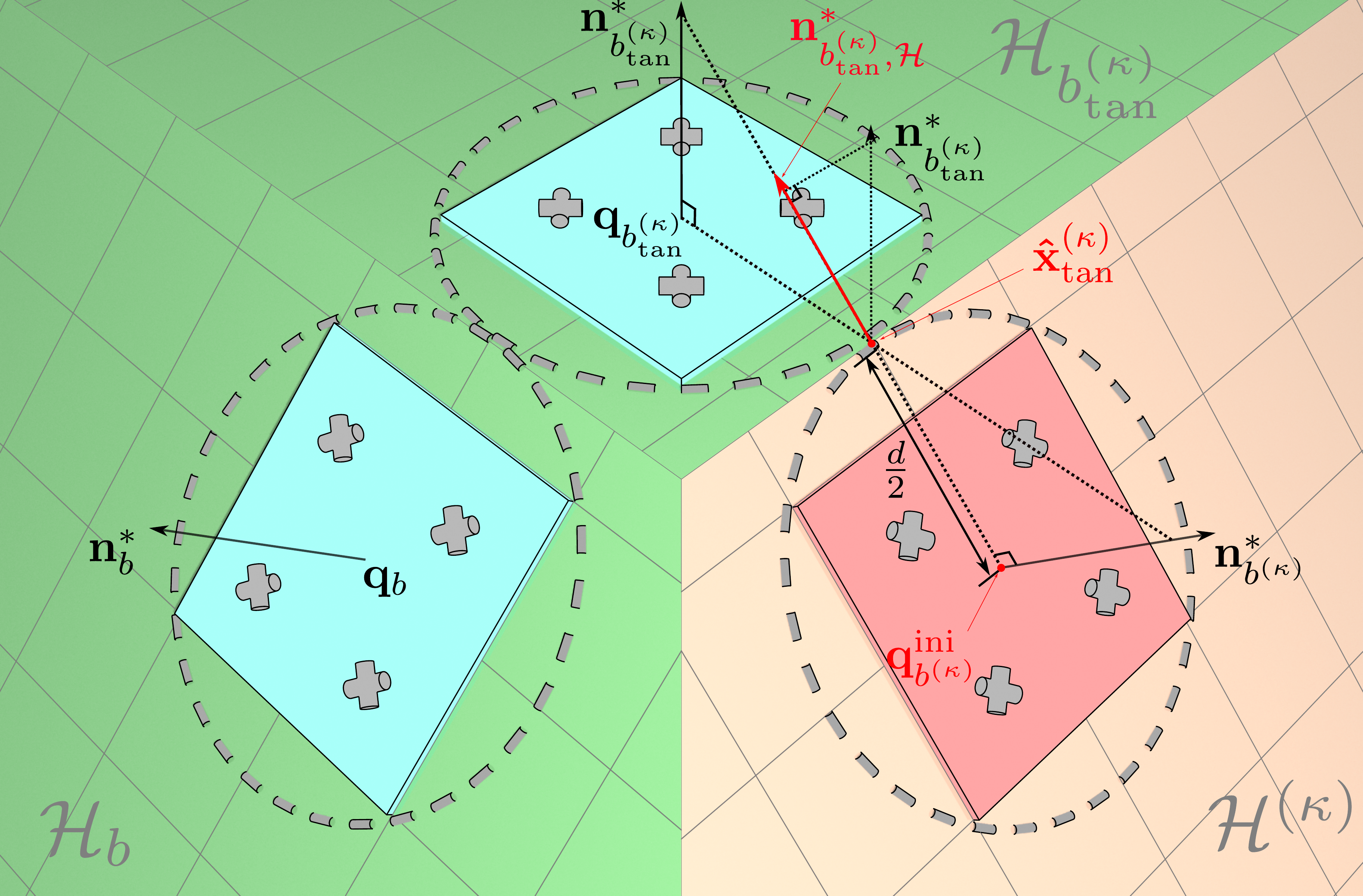}  
	\caption{Illustration for positioning the $b^{(\kappa)}$-th 6DMA surface in the $\kappa$-th iteration. The already positioned two 6DMA surfaces, indexed by $b,b_{\text{tan}}^{(\kappa)}\in\mc{X}^{(\kappa-1)}$, are located on the green hyperplanes, $\mc{H}_b$ and $\mc{H}_{b_{\text{tan}}^{(\kappa)}}$, respectively. The $b^{(\kappa)}$-th surface is to be positioned on the  orange hyperplane $\mc{H}^{(\kappa)}$, which is given by (\ref{get_plane1}) and (\ref{Compute_r}).  Following the proposed  positioning algorithm, the $b^{(\kappa)}$-th surface is placed at $\mbf{q}^{\text{ini}}_{b^{(\kappa)}}$, which is highlighted in red.}  
	\label{fig:intersection_point}  
\end{figure}

\item{Step 2:} Select the position of the $b^{(\kappa)}$-th 6DMA surface. Since the $b^{(\kappa)}$-th surface is placed on $\mc{H}^{(\kappa)}$, the CERs associated with 6DMA surfaces  in $\mc{X}^{(\kappa-1)}$ are already  within its non-positive halfspace. In this step, we  select $\mbf{q}_{b^{(\kappa)}}$ such that $	\mc{V}^{\text{e}}_{b^{(\kappa)}}(\mbf{q}_{b^{(\kappa)}})$ is also within the non-positive halfspaces of these CERs. Denote the projection of the  normal vector of the  $b^{(\kappa)}_{\text{tan}}$-th 6DMA surface (which is tangent to the hyperplane $\mc{H}^{(\kappa)}$) onto  $\mc{H}^{(\kappa)}$, as
\begin{align}
	\mbf{n}_{b^{(\kappa)}_{\text{tan}},\mc{H}}^*\triangleq \mbf{n}_{b^{(\kappa)}_{\text{tan}}}^* - \left({\mbf{n}^{*^\top}_{b^{(\kappa)}_{\text{tan}}}} \mbf{n}^*_{b^{(\kappa)}} \right)\mbf{n}^*_{b^{(\kappa)}}.
\end{align} 
Then, we obtain an initial guess for the position of the  $b^{(\kappa)}$-th 6DMA surface given by
\begin{align}
	\mbf{q}_{b^{(\kappa)}}^{\text{ini}} = \mbf{x}_{\text{tan}}^{(\kappa)} - \frac{d}{2}\frac{\mbf{n}_{b^{(\kappa)}_{\text{tan}},\mc{H}}^*}{\|\mbf{n}_{b^{(\kappa)}_{\text{tan}},\mc{H}}^*\|_2}. \label{q_ini_def}
\end{align}
 If $\mc{V}^{\text{e}}_{b^{(\kappa)}}(\mbf{q}_{b^{(\kappa)}}^{\text{ini}})$ lies within the non-positive halfspaces of the CERs associated with 6DMA surfaces  in $\mc{X}^{(\kappa-1)}$, we assign
\begin{align}
		\mbf{q}_{b^{(\kappa)}}  &\leftarrow	\mbf{q}_{b^{(\kappa)}}^{\text{ini}}.
\end{align}
Otherwise, we shift each already positioned surface indexed by $\mc{X}^{(\kappa-1)}$ outward along its associated normal vector projected onto $\mc{H}^{(\kappa)}$ by a distance $d/2$ in order to make more space for positioning the $b^{(\kappa)}$-th 6DMA surface, i.e., 
\begin{align}
	\mbf{q}_b  \leftarrow	\mbf{q}_b + \frac{d}{2}\frac{\mbf{n}^*_{b,\mc{H}}}{\|\mbf{n}^*_{b,\mc{H}}\|_2},~\forall b\in \mc{X}^{(\kappa-1)}, \label{q_update}
\end{align}
where the projected normal vector is given by
\begin{align}
 \mbf{n}^*_{b,\mc{H}} \triangleq \mbf{n}_b^* - \left(\mbf{n}^{*\top}_b \mbf{n}^*_{b^{(\kappa)}} \right)\mbf{n}^*_{b^{(\kappa)}}.
 \end{align}
Then, we assign
\begin{align}
	\mbf{q}_{b^{(\kappa)}}  &\leftarrow	\mbf{x}_{\text{tan}}^{(\kappa)}. \label{q_update_2}
\end{align}

\newtheorem{lemma}{Lemma}

\begin{lemma}
	 The 6DMA surfaces indexed by  $\{b^{(\kappa)}\} \cup \mc{X}^{(\kappa-1)}$  always satisfy the constraint (\ref{P_feasible_c1_new})  after the adjustment in (\ref{q_update}) and (\ref{q_update_2}).
	 \begin{proof}
	 	See Appendix A.
	 \end{proof}
\end{lemma}

\item{Step 3:} If $\kappa < B$, we update 
\begin{align}
	\mc{X}^{(\kappa)} &\leftarrow  \mc{X}^{(\kappa-1)} \cup \{b^{(\kappa)}\} ,\\
	\mc{Y}^{(\kappa)} &\leftarrow  \mc{Y}^{(\kappa-1)} \setminus \{b^{(\kappa)}\},\\
	\kappa &\leftarrow \kappa+1,
\end{align}
where $\setminus$ denotes the set difference operator.
We then proceed with the next iteration; otherwise, we stop the algorithm and return the final positions of the $B$ 6DMA surfaces, i.e., $\{\mbf{q}_b\}_{b=1}^B$.
 \end{itemize}

\begin{remark}
	The optimized positions  generally satisfy the constraint in (\ref{P_feasible_c3}). This holds practically, as the size of each 6DMA surface is typically much smaller than that of  \(\mathcal{V}_{\text{6DMA}}\). Specifically, if the edge length of \(\mathcal{V}_{\text{6DMA}}\)  (assumed to be a cube) is
	\begin{align}
		d_{\text{max}} = B d, \label{bound}
	\end{align}
	then \(B\) 6DMA surfaces, each with an arbitrary rotation, can be placed within \(\mathcal{V}_{\text{6DMA}}\) using our proposed positioning algorithm,  since the increase in the distance between the two farthest points among the positioned 6DMA surfaces does not exceed \(d\) in each iteration. It is worth noting that the bound in (\ref{bound}) is conservative and  rarely reached with the positions obtained by our proposed algorithm with given rotations, which usually results in a much smaller edge length of $\mc{V}_{\text{6DMA}}$ than $d_{\text{max}}$ given in (\ref{bound}), as shown by the numerical results in the next section.
\end{remark}

\section{Simulation Results}

\begin{figure}[!t]  
	\centering
	\includegraphics[width=2.5in]{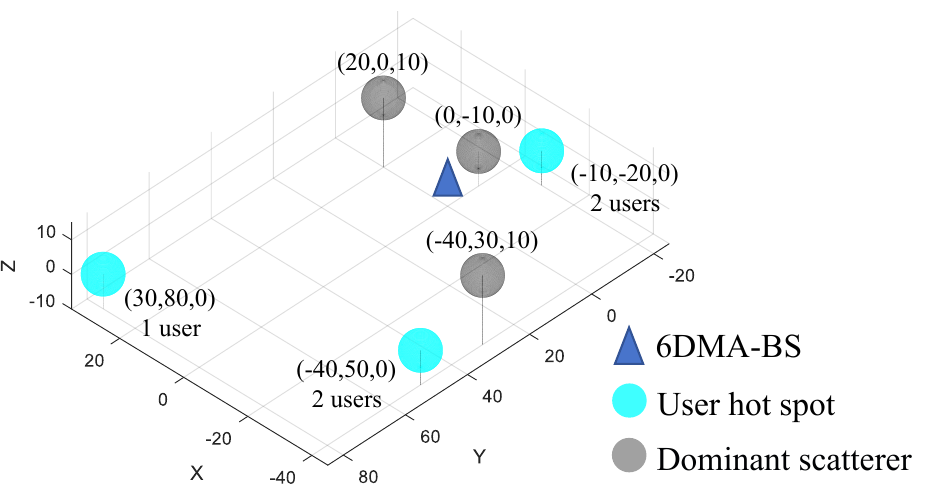}  
	\caption{Simulation setup.}  
	\label{fig:simulation_setup}  
\end{figure}

\begin{figure}[!t]  
	\centering
	\includegraphics[width=1.65in]{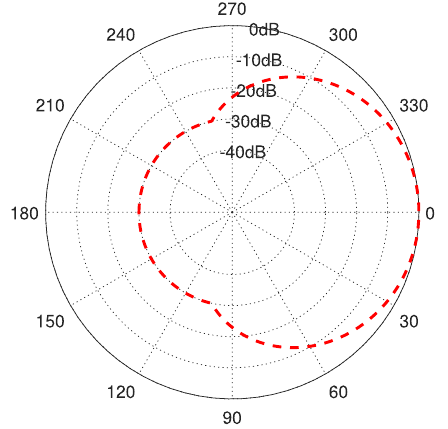}  
	\caption{Antenna horizontal radiation pattern in simulation.}  
	\label{fig:antenna_pattern}  
\end{figure}

This section presents simulation results to validate our proposed schemes for SCI estimation and position-rotation optimization based on SCI in 6DMA-enabled communication systems.  In the simulation, we consider three user clusters and three dominant scatterers. As illustrated in Fig.\,\ref{fig:simulation_setup}, the 6DMA-BS is located at \((0, 0, 0)\), and the positions of the dominant scatterers are \((-40, 30, 10)\), \((20, 0, 10)\), and \((0, -10, 0)\), respectively, with all coordinates given in meter (m). The user distribution is as follows: two users are randomly located within a sphere of radius \( r_1 = 5\,\text{m} \) centered at \((-40, 50, 0)\); one user is randomly located within a sphere of radius \( r_2 = 5\,\mathrm{m} \) centered at \((30, 80, 0)\); and two users are randomly located within a sphere of radius \( r_3 = 10\,\mathrm{m} \) centered at \((-10, -20, 0)\). The path loss model is given by \( v = \left(\frac{\lambda}{4\pi}\right)^2 d^{-\eta} \), where the path loss exponent is \( \eta = 3 \), and the carrier wavelength is \( \lambda = 0.125\,\mathrm{m} \). The direct links between all users and the BS are assumed to be blocked, and all users share the same set of dominant scatterers. 
 Each 6DMA surface consists of \( N = 4 \) antennas, as illustrated in Fig.\,\ref{fig:single_surface}. The edge length of each 6DMA (square) surface is \(  \lambda \), and the positions of its four antennas, \( \bar{\mathbf{r}}_n \)'s, are located at \( [0, \pm \frac{1}{4}\lambda, \pm \frac{1}{4}\lambda] \) in its local coordinate system. The edge length of the cubic region where the $B$ 6DMA surfaces can be flexibly positioned/rotated,  \( \mathcal{V}_{\text{6DMA}} \), is set to \( 1\,\text{m} \). The antenna gain pattern in (\ref{antenna_gain}) follows the 3GPP standard \cite{3gpp2017}, with the horizontal radiation pattern shown in Fig.\,\ref{fig:antenna_pattern}.

Unless otherwise stated, the 6DMA-BS is equipped with \( B = 8 \) 6DMA surfaces, and the main-lobe beamwidth of each antenna is set to \( \Theta = 65^\circ \).  In Stage~I of the proposed protocol, the number of training position–rotation pairs used for channel measurements is \( M = 16 \), and the number of channel realizations used for time averaging is \( T = 100 \). For the OMP algorithm to solve (P2), the discrete DoA grid consists of \( I = 360 \times 180 \) directions. In the algorithm in Section~V-A for rotation optimization, the number of candidates for greedy search is \( \bar{M} = 512 \), the maximum number of iterations is \( \kappa_{\max} = 20 \), and the step size for the gradient approximation in (\ref{Nabla_1}) is set to \( \epsilon = 2^{-16} \).

For performance comparison, we consider the following benchmark schemes.
\begin{enumerate}

 \item Monte Carlo with Alternating Optimization (MC-AO):   the expectation over $r_k$ is computed by the Monte Carlo method as shown in (\ref{r_MC}), instead of using its approximation $\underline{r}_k$. 
 In this  scheme, the distribution of $\mbf{H}$ is assumed to be perfectly known, and the  number of channel realizations  is set to  $W=10^3$. In addition, the alternating optimization method proposed in \cite{6DMA_TWC} is used to optimize the 6DMA positions and rotations. 
\item FA-BS:  a fixed three-sector BS is considered, with each sector consisting of 11 fixed (position and rotation)  antennas (FAs). The elevation tilt angle is set as 0. 
\item PAA-BS: a three-sector BS is considered, where each sector is equipped with 11 position-adjustable antennas (PAAs), which can move freely on each sector's two-dimensional (2D) surface with size $0.5\,\mathrm{m}\times0.5\,\mathrm{m}$ and inter-antenna minimum spacing equal to $\lambda/2$. The antenna positions are optimized using the particle swarm optimization (PSO) algorithm\cite{PSO} based on users' instantaneous channels.

\end{enumerate} 

\begin{figure}[!t]  
	\centering
	\includegraphics[width=2.55in]{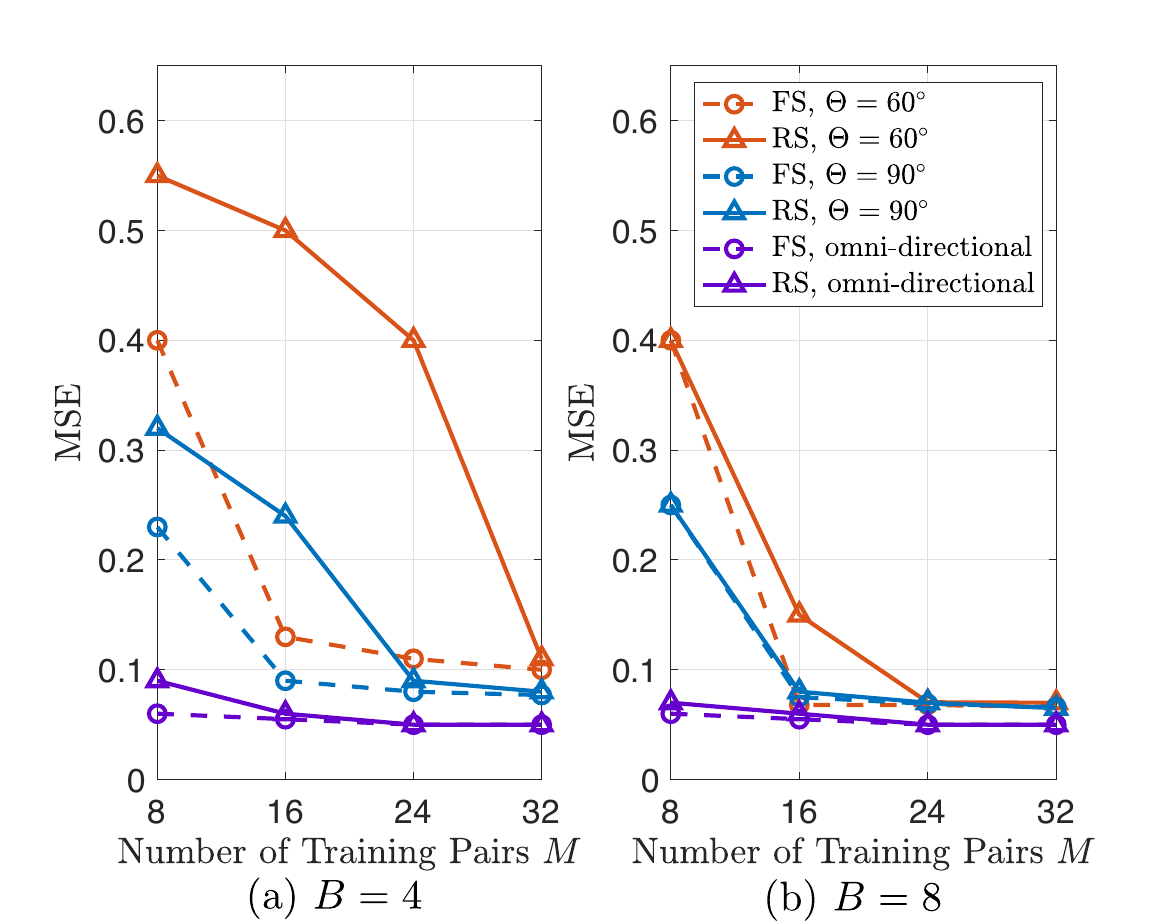}  
	\caption{SCI estimation MSE versus number of training position-rotation pairs for channel measurement.}  
	\label{result_MPC}  
\end{figure}

First, Fig.\,\ref{result_MPC} illustrates the impact of the number of training position–rotation pairs \( M \), antenna beamwidth \( \Theta \), and the number of 6DMA surfaces \( B \) on the normalized SCI estimation mean squared error (MSE), denoted by
\begin{align}
\frac{\|\mbf{\Sigma}^{\text{per}} - \mbf{\Sigma}^{\text{est}}\|_F}{\|\mbf{\Sigma}^{\text{per}}\|_F + \|\mbf{\Sigma}^{\text{est}}\|_F },
\end{align}
where $\mbf{\Sigma}^{\text{true}}$ and $\mbf{\Sigma}^{\text{est}}$ denote the perfect SCI and estimated SCI based on the $M$ training position-rotation pairs, respectively. We consider a full-sampled (FS) method  for performance comparison with our proposed RS method with a smaller number of 6DMA surfaces for channel measurements. Specifically, for the FS method, we assume that  there are $B = M$  6DMA surfaces  deployed, which can simultaneously measure the channels at the $M$ 6DMA training position-rotation pairs.  In accordance with~(\ref{M_require}), it is observed that both Fig.\,\ref{result_MPC}(a) and \ref{result_MPC}(b) show that narrower antenna beam requires a larger \( M \) to sufficiently cover the spatial domain for accurately resolving the MPCs.  Moreover, by comparing the MSEs shown in Figs.\,\ref{result_MPC}(a) and \ref{result_MPC}(b),  it is observed that as   \( B \) increases, the performance of the proposed RS method approaches that of the FS benchmark,  at the expense of more 6DMA surfaces deployed.

\begin{figure}[!t]  
	\centering
	\includegraphics[width=2.55in]{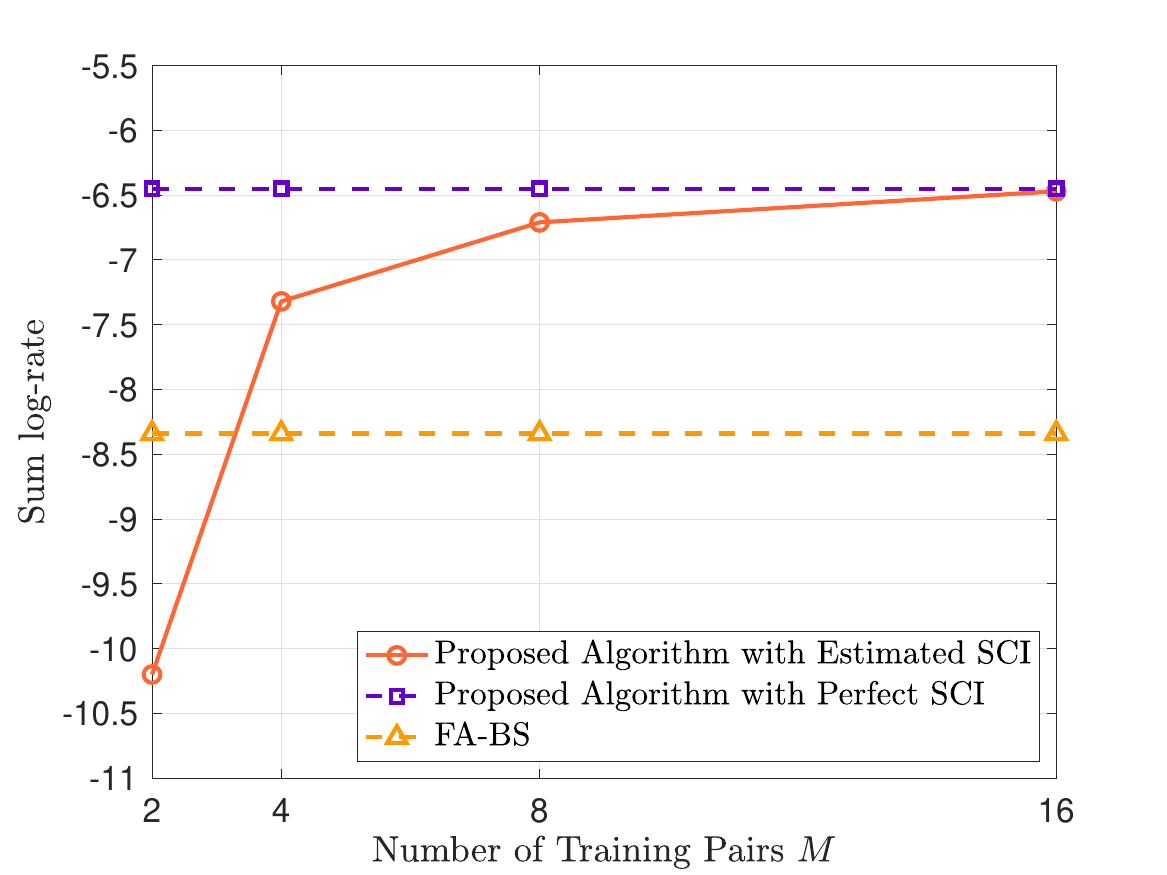}  
	\caption{Network  average sum log-rate versus number of training position-rotation pairs for channel measurement.}  
	\label{result_cvm}  
\end{figure}

Fig.\,\ref{result_cvm} illustrates the network average sum log-rate versus the number of training position–rotation pairs \( M \)  for  channel measurements. As shown in the figure, the performance of  the proposed 6DMA position and rotation optimization  algorithm with estimated SCI improves rapidly as \( M \) increases due to more accurate estimation of SCI  using more channel measurements. The sum log-rate gradually saturates for sufficiently large  \( M \), indicating that sufficient channel measurements have been taken to estimate the SCI accurately and achieve  near-optimal 6DMA positioning and rotation based on estimated SCI. Notably, the proposed algorithm with estimated SCI approaches the performance of the case with perfect SCI as \( M \) increases, demonstrating the effectiveness of the estimated SCI based position-rotation optimization strategy.

\begin{figure}[!t]  
	\centering
	\includegraphics[width=2.55in]{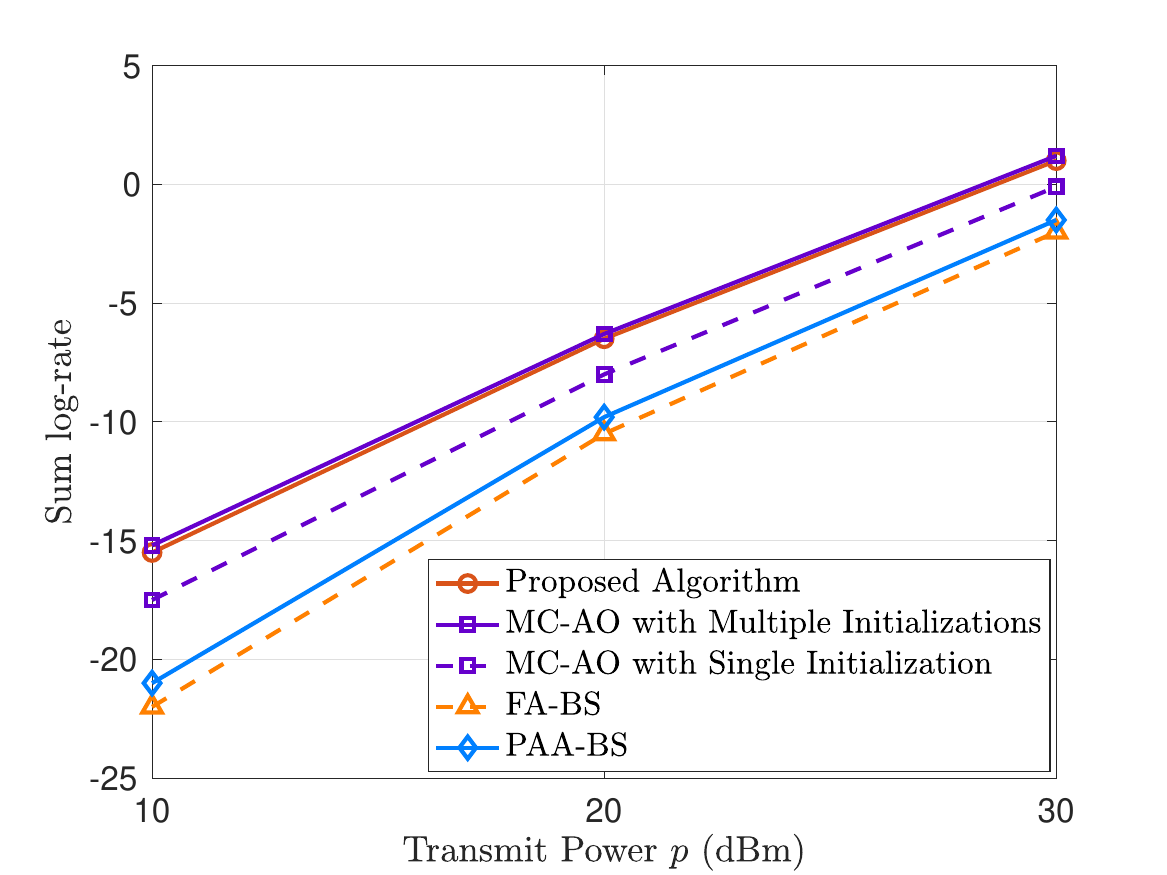}  
	\caption{Network  average sum log-rate versus transmit power  of users.}  
	\label{result_cvp}  
\end{figure}

In Fig.\,\ref{result_cvp}, we evaluate the network average sum log-rate   versus  user transmit power. For simplicity,  equal transmit power is assumed for all users, i.e., $p_k = p$, $\forall k$. Among the compared schemes, the MC-AO method with multiple initializations achieves the best performance by accurately approximating the expectation of the sum log-rate given in (\ref{r_origin}). However, the MC-AO method has  two major drawbacks compared to our proposed sequential rotation and position optimization: firstly, it entails substantially higher computational complexity due to excessive channel sampling; secondly, it is prone to poor local optima due to the use of alternating optimization over positions and rotations, thus requiring multiple random initializations to achieve good performance (as shown by the performance gap for this scheme with single initialization versus multiple initializations). In contrast, our proposed algorithm achieves comparable performance with much lower computational complexity, demonstrating superior capability for balancing performance and complexity. The PAA method slightly outperforms the baseline FA due to  adjustability of antenna positions only, but performs significantly inferior to the proposed algorithm, which can fully exploit the antenna position and rotation adjustments based on users' SCI. 

\begin{figure}[!t]  
 	\centering
 	\includegraphics[width=2.55in]{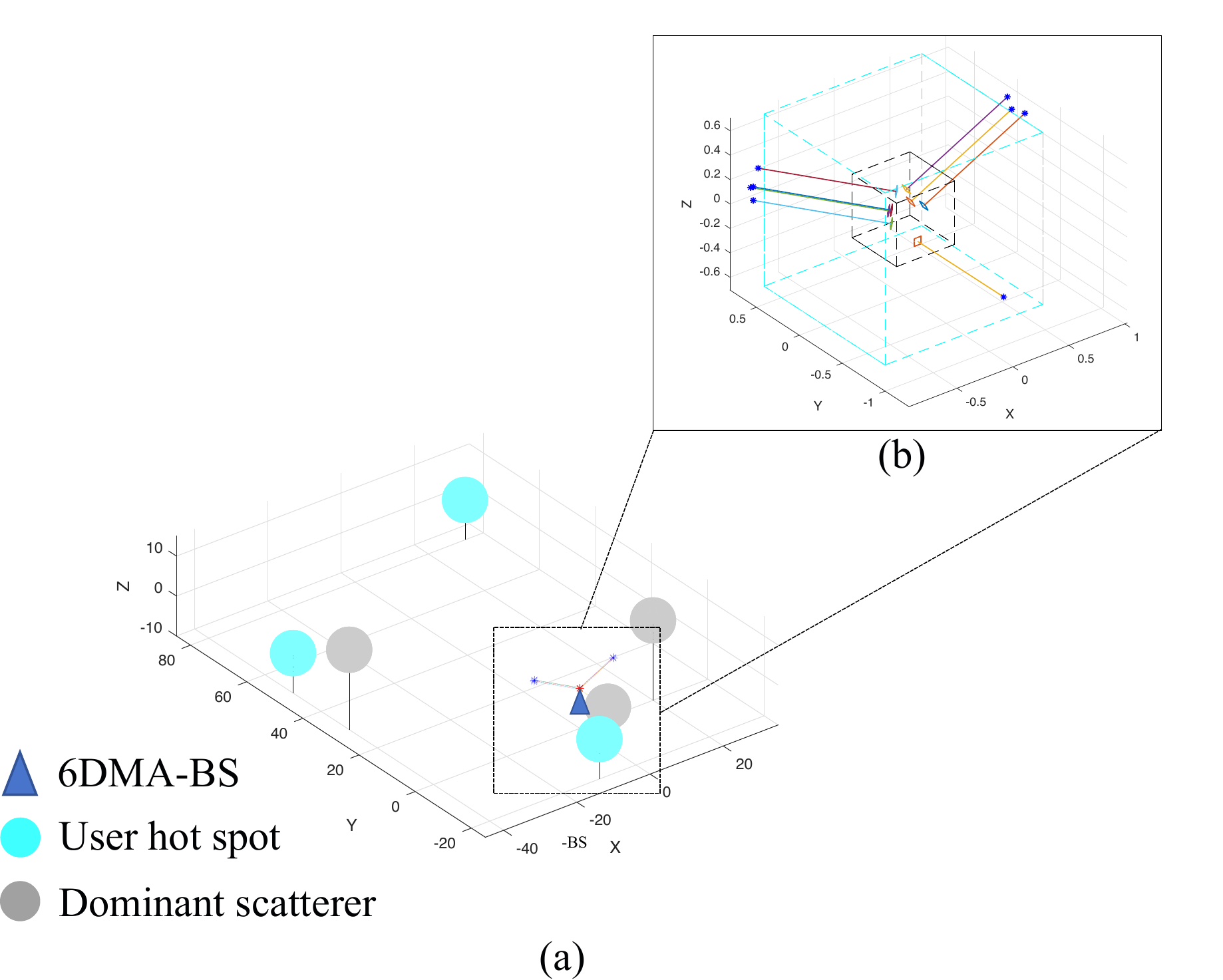}  
 	\caption{Illustration of the positions and rotations  of 6DMA surfaces optimized by the proposed algorithm.}  
 	\label{result_rp}  
 \end{figure}

 Finally,  Fig.\,\ref{result_rp} illustrates the  optimized positions and rotations of 6DMA surfaces  by the proposed algorithm. Specifically, Fig.\,\ref{result_rp}(a) visualizes the orientations of the 6DMA surfaces in the global coordinate system, with solid lines and
 star markers indicating their optimized normal vectors, while Fig.\,\ref{result_rp}(b) shows the optimized surface positions and rotations in the 6DMA-BS region. In Fig.\,\ref{result_rp}(b), all  6DMA surfaces are placed within a cube (indicated by black dash lines) with  edge length $d' = 0.52\,\mathrm{m}$, which is much smaller than the  bound $d_{\text{max}} = Bd = 1.41\,\mathrm{m}$  (indicated by cyan dash lines)  given in (\ref{bound}), indicating the effectiveness of the proposed method for finding the feasible 6DMA surfaces' positions given their designed rotations within the 6DMA-BS region.  

\section{Conclusion}
This paper presents  a practical low-complexity design framework for 6DMA-enabled communication systems with SCI estimation and SCI-based 6DMA position and rotation optimization.  We first propose a practical protocol for 6DMA-BS and  an OMP-based method for the reconstruction of SCI of all users, by resolving the MPC information including  the DOAs and average power values of all channel MPCs. An optimization problem is then formulated to maximize the average sum log-rate of a multi-user MAC based on estimated SCI. To reduce computational complexity and improve performance over the existing approach based on MC approximation and AO, we propose a new sequential optimization method that first determines 6DMA rotations and then finds their feasible positions to realize the optimized rotations subject to practical antenna placement constraints.  Simulation results demonstrate that the proposed algorithm  based on estimated SCI  achieves performance close to the MC-AO benchmark with perfect knowledge of SCI, but with significantly lower computational complexity. Furthermore, the results reveal  that the rotation adjustment of 6DMA
surfaces plays a more significant role in enhancing performance over traditional BS with FAs, as  compared to position adjustment, thus providing new insight  into the performance enhancement mechanism
of 6DMA systems.

\section*{Appendix: Proof of Lemma 1}

	Consider two arbitrary surfaces indexed by \( b,~b' \in \mathcal{X}^{(\kappa-1)} \),  \( b \ne b' \). Let \( \mathbf{x} \) be an arbitrary point in \( \mathcal{V}^{\text{e}}_{b'}(\mathbf{q}_{b'}) \). After the shift, we have
	\begin{align}
		&\mathbf{n}_b^{*\top} \left( \left( \mathbf{x} + \frac{d}{2} \frac{\mathbf{n}^*_{b',\mathcal{H}}}{\| \mathbf{n}^*_{b',\mathcal{H}} \|_2} \right) - \left( \mathbf{q}_b + \frac{d}{2} \frac{\mathbf{n}^*_{b,\mathcal{H}}}{\| \mathbf{n}^*_{b,\mathcal{H}} \|_2} \right) \right) \nonumber \\
		&\overset{(a)}{\le} \mathbf{n}_b^{*\top} \left( \frac{d}{2} \frac{\mathbf{n}^*_{b',\mathcal{H}}}{\| \mathbf{n}^*_{b',\mathcal{H}} \|_2} - \frac{d}{2} \frac{\mathbf{n}^*_{b,\mathcal{H}}}{\| \mathbf{n}^*_{b,\mathcal{H}} \|_2} \right) \nonumber \\
		&\overset{(b)}{=} \mathbf{n}^*_{b,\mathcal{H}}{}^{\top} \left( \frac{d}{2} \frac{\mathbf{n}^*_{b',\mathcal{H}}}{\| \mathbf{n}^*_{b',\mathcal{H}} \|_2} - \frac{d}{2} \frac{\mathbf{n}^*_{b,\mathcal{H}}}{\| \mathbf{n}^*_{b,\mathcal{H}} \|_2} \right) \nonumber \\
		&\overset{(c)}{\le} 0,
	\end{align}
	where $(a)$ holds because \( \mathbf{x} \in \mathcal{V}^{\text{e}}_{b'}(\mathbf{q}_{b'}) \subseteq \mathcal{V}_b(\mathbf{q}_b) = \{ \mathbf{x} \in \mathbb{R}^3 \mid \mathbf{n}_b^{*\top} (\mathbf{x} - \mathbf{q}_b) \le 0 \} \), and thus \( \mathbf{n}_b^{*\top} (\mathbf{x} - \mathbf{q}_b) \le 0 \);  
	$(b)$ holds because the component of \( \mathbf{n}_b^* \) orthogonal to \( \mathcal{H}^{(\kappa)} \) is also orthogonal to both \( \mathbf{n}^*_{b,\mathcal{H}} \) and \( \mathbf{n}^*_{b',\mathcal{H}} \), yielding zero inner-product;  
	$(c)$ follows from the Cauchy–Schwarz inequality. Therefore, constraint~(\ref{P_feasible_c1_new}) is satisfied for the surfaces indexed by \( \mathcal{X}^{(\kappa-1)} \) after being shifted.
	
	On the other hand, before the shift, we have
	\begin{align}
		\mathbf{x}_{\text{tan}}^{(\kappa)} \in \mathcal{H}^{(\kappa)} \cap \left( \bigcap_{b \in \mathcal{X}^{(\kappa-1)}} \mathcal{H}_b^-(\mathbf{q}_b, \mathbf{u}_b^*) \right) \triangleq \mathcal{P}.
	\end{align}
	After the shift, the boundaries of \( \mathcal{P} \) move outward along their respective normal vectors by a distance \( d/2 \), forming a new region \( \mathcal{P}^{\text{new}} \) given by
	\begin{align}
		\mathcal{P}^{\text{new}} = \mathcal{H}^{(\kappa)} \cap \left( \bigcap_{b \in \mathcal{X}^{(\kappa-1)}} \mathcal{H}_b^-\left( \mathbf{q}_b + \frac{d}{2} \frac{\mathbf{n}^*_{b,\mathcal{H}}}{\| \mathbf{n}^*_{b,\mathcal{H}} \|_2}, \mathbf{u}_b^* \right) \right). \label{CCCx}
	\end{align}
	Since the distance from \( \mathbf{x}_{\text{tan}}^{(\kappa)} \) to all the boundaries of \( \mathcal{P}^{\text{new}} \) is no less than \( d/2 \), the CER \( \mathcal{V}^{\text{e}}_{b^{(\kappa)}}(\mathbf{x}_{\text{tan}}^{(\kappa)}) \), centered at \( \mathbf{x}_{\text{tan}}^{(\kappa)} \) with radius \( d/2 \), is also contained within \( \mathcal{P}^{\text{new}} \). Thus,
	\begin{align}
		\mathcal{V}^{\text{e}}_{b^{(\kappa)}}(\mathbf{q}_{b^{(\kappa)}}) \subseteq \mathcal{P}^{\text{new}}, \label{CCC}
	\end{align}
	where \( \mathbf{q}_{b^{(\kappa)}} = \mathbf{x}_{\text{tan}}^{(\kappa)} \) is updated according to~(\ref{q_update_2}). Relation~(\ref{CCC}) implies that
	i) all CERs associated with the surfaces indexed by \( \mathcal{X}^{(\kappa-1)} \) lie within the non-positive halfspace of the \( b^{(\kappa)} \)-th surface since the \( b^{(\kappa)} \)-th surface lies on \( \mathcal{H}^{(\kappa)} \);  
	and ii) the CER of the \( b^{(\kappa)} \)-th surface lies within the non-positive halfspaces of all surfaces in \( \mathcal{X}^{(\kappa-1)} \) with the updated positions.  
	
	Hence, constraint~(\ref{P_feasible_c1_new}) is satisfied for the surfaces indexed by both  $\mc{X}^{(\kappa-1)} $ and $\{b^{(\kappa)}\}$ after the adjustments in~(\ref{q_update}) and~(\ref{q_update_2}). Lemma 1 is thus proved.

\normalem
\bibliographystyle{IEEEtran}
\bibliography{6DMA_bib}

\end{document}